\pdfoutput=1
\documentclass[reprint,aip,jcp,twocolumn,showpacs]{revtex4-1}
\usepackage[paperwidth=210mm,paperheight=297mm,centering,hmargin=2.0cm,vmargin=2.5cm]{geometry}
\usepackage{soul}
\usepackage[utf8]{inputenc}
\usepackage[normalem]{ulem}
\usepackage{mathptmx}
\usepackage{mathtools}
\usepackage{amsmath} 
\usepackage{braket}
\usepackage{float}
\DeclarePairedDelimiter\abs{\lvert}{\rvert}%
\usepackage{lipsum}  
\usepackage{svg}

\usepackage{mathptmx}
\usepackage{stackrel}
\usepackage{mathtools}
\usepackage{amsmath} 
\usepackage{braket}
\usepackage{float}
\usepackage{lipsum}  
\usepackage{svg}

\renewcommand\thesection{\arabic{section}}
\usepackage[sort&compress]{natbib}
\usepackage{mhchem}
\usepackage{changepage}
\usepackage{float}
\usepackage{mathrsfs}
\usepackage{tikz}
\usepackage{mathrsfs}
\usepackage{tikz}

\date{}

\begin{document}

\title{Gated reactions in discrete time and space}

\author{{\normalsize{}Yuval Scher}
{\normalsize{}}}
\email{yuvalscher@mail.tau.ac.il}

\author{{\normalsize{}Shlomi Reuveni}
{\normalsize{}}}
\email{shlomire@tauex.tau.ac.il}

\affiliation{\noindent \textit{$^{1}$School of Chemistry, The Center for Physics and Chemistry of Living Systems, The Raymond and Beverly Sackler Center for Computational Molecular and Materials Science, \& The Mark Ratner Institute for Single Molecule Chemistry, Tel Aviv University, Tel Aviv 6997801, Israel}}

\date{\today}


\begin{abstract}

\noindent How much time does it take two molecules to react? If a reaction occurs upon contact, the answer to this question boils down to the classic first-passage time problem: find the time it takes the two molecules to meet. However, this is not always the case as molecules switch stochastically between reactive and non-reactive states. The reaction is then said to be ``gated'' by the internal states of the molecules involved which could have a dramatic influence on kinetics. A unified, continuous-time, approach to gated reactions on networks was presented in [Phys. Rev. Lett. 127, 018301, 2021]. Here, we build on this recent advancement and develop an analogous discrete-time version of the theory. Similar to continuous-time, we employ a renewal approach to show that the gated reaction time can always be expressed in terms of the corresponding ungated first-passage and return times; which yields  formulas for the generating function of  the  gated  reaction-time distribution and its corresponding mean and variance. In cases where the mean reaction time diverges, we show that the long-time asymptotics of the gated problem is inherited from its ungated counterpart. However, when molecules spend most of their time  non-reactive, an interim regime of slower power-law decay emerges prior to the terminal asymptotics. The discretization of time also gives rise to resonances and anti-resonances, which were absent from the continuous time picture. These features are illustrated using two case studies that also demonstrate how the general approach presented herein greatly simplifies the analysis of gated reactions. 

\end{abstract}

\maketitle

\section*{I. Introduction}

\noindent Gated processes are ubiquitous in the chemical, \cite{szabo1982stochastically,berezhkovskii1997smoluchowski,makhnovskii1998stochastic,bandyopadhyay2000theoretical,benichou2000kinetics} biological \cite{mccammon1981gated,reingruber2009gated,boehr2009role, changeux2011conformational,vogt2012conformational} and physical sciences. \cite{budde1995transient,caceres1995theory,re1996survival,spouge1996single,sheu1997survival,sheu1999first,kumar2021first} An example of a gated process is given in Fig. 1, which illustrates a gated cleavage reaction. Adopting a chemical prone nomenclature, we imagine a ``particle", e.g., enzyme, that can react with a ``target", e.g., a cleavage site on a peptide or DNA strand. The particle switches between two states -- a reactive state and a non-reactive state. Importantly, a reaction occurs only when the particle meets the target while being in the reactive state. Thus, the reaction is said to be ``gated" by the particle's internal state.    

More generally, a gated reaction has an underlying spatial component on top of which an internal (gating) component is added. The spatial component can be any stochastic motion that the particle undertakes in search of the target. \cite{redner2001guide,metzler2014first,klafter2011first,rudnick2004elements} Similarly, one can in principal consider any internal dynamics, e.g, multiple reactive and/or non-reactive states;\cite{spouge1996single} or even non-Markovian dynamics.\cite{caceres1995theory}

Recently, gated processes enjoyed renewed attention across a wide range of disciplines and directions: Kochugaeva, Shvets and Kolomeisky considered the switching dynamics of a protein undertaking facilitated diffusion on a DNA strand.\cite{kochugaeva2016conformational} A later general model, that bears some similarity, is the gated continuous-time and discrete-space random search for a target on a 1D interval by Shin and Kolomeisky.\cite{shin2018molecular} Markedly, both models allow for diffusion rates which depend on the internal state. Gated diffusion with different diffusivities was also considered by Godec and Metzler, for 3D Brownian motion inside a spherical domain, with a spherical target at its center, \cite{godec2017first} thus generalizing the pioneering works of Szabo et al.\cite{szabo1982stochastically,szabo1980first} 

\begin{figure}[t]
\begin{centering}
\includegraphics[width=1\linewidth]{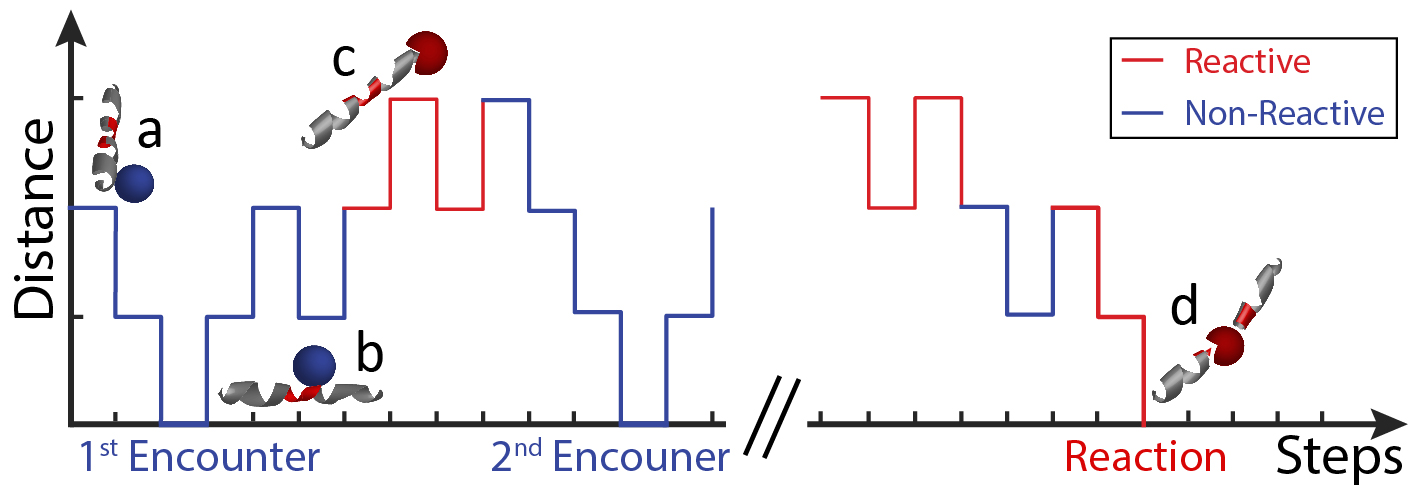}
\caption{An illustration of a gated reaction in discrete time and space. A particle, here a cleaving enzyme, switches between a reactive state (red) and a non-reactive state (blue). A cleaving reaction occurs when the enzyme reaches the cleavage site (target), which is denoted here by a red patch on a gray peptide, while being in the reactive state. Different stages of the reaction were denoted as follows: (a) The enzyme is away from the cleavage site and in the non-reactive state, (b) The enzyme is on the cleavage site and in the non-reactive state, (c) The enzyme is away from the cleavage site and in the reactive state, (d) The enzyme is on the cleavage site and in the reactive state, which results in a reaction.}
\end{centering}
\end{figure} 

Even when considering a particle with diffusivity that is decoupled from the internal state there are still plenty directions to explore. Mercado-Vásquez and Boyer studied the reaction time of a gated 1D diffusing particle on the semi-infinite line,\cite{mercado2019first} adding novel insights to the works of Budde, Cáceres and Ré.\cite{budde1995transient,caceres1995theory,re1996survival} They then proceeded to extend their theory to the cases of a gated run-and-tumble particle \cite{mercado2021first} and a gated diffusive particle under restart.\cite{mercado2021search} The latter was also previously considered by Bressloff with a slightly different model.\cite{bressloff2020diffusive} Gopic and Szabo deepened the exploration of multiple gated particles/targets in a model that allows for  reversible binding.\cite{gopich2016reversible} Lawley and Keener demonstrated mathematically the connection between the Robin (radiation) boundary condition, which is central in the theory of diffusion-influenced reactions,\cite{calef1983diffusion,weiss1986overview} and a gated boundary.\cite{lawley2015new}

In a previous letter this year, we developed a unifying approach to gated reactions on networks in the framework of the Montroll-Weiss Continuous-Time Random-Walk (CTRW).\cite{scher2020unifying} We showed that the mean and distribution of the gated reaction time can always be expressed in terms of ungated first-passage and return times. A related result is that of Spouge, Szabo and Weiss, that connected between gated and ungated propagators of Markovian diffusion processes.\cite{spouge1996single} We instead focused on the reaction time distribution, which is generally more attainable, be it analytically, experimentally or numerically. Utilizing a renewal framework, we were able to treat general spatial processes, which can be non-homogenous and non-Markovian.\cite{scher2020unifying} Here, we develop an analogous discrete-time version of this framework, and illustrate its applicability with examples.

Previously, a discrete time multi-particle gating reaction on a d-dimensional regular lattice was considered by Bénichou, Moreau and Oshanin.\cite{benichou2000kinetics} However, to the best of our knowledge, \textit{single-particle} discrete time and space gating problems were not considered before. One reason to do so is because discrete space-time models are the bread-and-butter of stochastic analysis and first-passage modelling.\cite{redner2001guide,klafter2011first,rudnick2004elements,Gallager2013book} Discrete-time Markov chains in particular are widely applied in this context, and extending these to include gating dynamics is thus desirable. Not less important is the fact that  discrete space-time models of gating provide a convenient testbed for resonance and anti-resonance effects that were completely absent from the class of continuous time models considered in our previous work.\cite{scher2020unifying} While such effects can still be studied in continuous time using models of non-Markovian internal gating dynamics, understanding their origin within a much simpler discrete-time Markovian framework has significant advantages in terms of analytical tractability.

The remainder of this paper is structured as follows. In section II, we develop a general theory for gated reactions on networks. Working in discrete time, we employ a renewal approach to express the gated reaction-time in terms of the corresponding ungated reaction-time. Formulas for the generating function of the gated reaction-time distribution, as well its mean and variance, are derived. In cases where the mean reaction time diverges, we prove that the long-time asymptotics of the gated problem is inherited from its ungated counterpart, where only the pre-factor of the power-law tail changes. We also study these cases under conditions of high-cripticity, i.e., when the molecules spend most of their time in the non-reactive state. We show that the terminal asymptotic regime is then preceded by a transient regime of slower power-law decay which we determine exactly.

In sections III and IV, we utilize the general results obtained to analyze two case studies. The first is of a particle moving in a unidirectional manner on a circle in search of a target. While the ungated version of this problem is nothing but trivial, it is more challenging to solve the gated version, and indeed interesting features arise when gating is added. Specifically, we see that while the results in section II bear resemblance to the results of Ref.  \cite{scher2020unifying}, the discretization of time introduces new features that do not exist in the continuous-time analogue. Noticeably, resonance and anti-resonance effects emerge, when the internal gating dynamics is in, or out of, phase with the spatial process governing particle-target encounters.

The case study considered in section III belongs to a wide class of problems in which the mean reaction time is well-defined. Yet, one often encounters problems in which the mean diverges, e.g., when considering open systems. In such cases, one is usually interested in the reaction time distribution itself, and especially in its long-time asymptotics. In section IV we demonstrate how our formalism can be used to deal with such cases by using it to solve for the gated reaction time of a symmetric random-walk on a 1D lattice. As this problem was not solved previously, we offer two approaches for the solution: (a) writing the corresponding master equation and solving it brute-force, (b) using the renewal approach developed in section II of this paper. One can then appreciate how the latter elegantly replaces a complex calculation in favour of a straight forward solution method which readily reveals key features of the gated analogue of a classic first-passage problem. To put things in context, we conclude section IV with a general perspective on gated 1D random walks and reactions. 

The \textit{modus operandi} employed in sections III and IV is not limited to 1D systems, and applies without change to gated reactions on lattices of higher dimensions and to gated reactions on networks of arbitrary topology. This fact is further stressed in section V which concludes this work with a simple ``algorithm'' that allows one to generate solutions to gated reaction problems from solutions to their ungated counterparts, thus circumventing the need to solve gated problems directly. Importantly, this approach can be utilized even when an analytical solution to the ungated first-passage problem is not known, and as long as it can be estimated numerically. In this way, the analysis of gated problems is greatly simplified.

\section*{II. General Theory}\label{General_Theory}

\noindent Consider a single particle in search of a single target site that is embedded in a general network (N.B., the network can be a lattice, but it need not be). To find the target, the particle conducts a general random walk on the network. This is done in discrete time. Namely, in each time step the particle jumps from its site of origin to a different site according to some prescribed law of motion. To keep things general, we allow for non-homogeneous random walks, where the set of accessible sites and the probabilities to reach each one of them can depend on the site of origin. Our only requirement is that the prescribed law of motion is not time dependent, i.e., that the connectivity at each site, and its jumping probabilities, do not change in time. We will refer to this property as time-homogeneity.

In contrast to classical first-passage problems,\cite{redner2001guide,metzler2014first,klafter2011first,rudnick2004elements} in gated problems the particle stochastically transitions between a reactive state and a non-reactive state, such that arrival to the target is not sufficient by itself. To react, the particle must reach the target in the reactive state. We are interested in the statistics of the reaction time, defined to be the first-passage to target while in the reactive state. Our theory yields close formulas for the mean, variance and the generating function of the probability mass function (PMF) of the reaction time in terms of the corresponding ungated first-passage statistics. It is important to note that we hereby focus on a single particle and a single target, and it is of no matter whether gating occurs on the particle, on the target, or is due to a joint effect. We will thus often remain general and consider ``a gated system". In the case of multiple particles, however, the statistics for gated particles and gated targets do differ.\cite{berezhkovskii1997smoluchowski,makhnovskii1998stochastic,gopich2016reversible,zhou1996theory} 

In what follows, we use $\langle X\rangle$, $\sigma(X)$ and $\tilde{X}(z) \equiv\left\langle z^{X}\right\rangle = \sum_{n=0}^{\infty} P(X=n) z^{n}$ to denote, respectively, the expectation, standard deviation, and Z-transform of a integer-valued random variable $X$.

\subsection*{Internal Dynamics of a Two-State System}
\noindent The internal dynamics of the particle is described by a discrete-time Markov chain composed of two states (Fig. 2): Reactive (R) and Non-Reactive (NR). The transition probability from R to NR is denoted by $p$ and the complementary self-transition probability is denoted by $1-p$. Similarly, the transition probability from NR to R is denoted by $q$ and the complementary self-transition probability is denoted by $1-q$.

We are interested in $\text{P}(\text{R},n \mid  \omega_{0})$, the conditional probability to be in the reactive state after $n$ steps, given an initial internal state $ \omega_{0} \in \{\text{R, NR, eq}\}$, where `eq' stands for equilibrium. Solving, we obtain (Appendix A)

\begin{equation}  \label{eq:1}
\begin{array}{ll}
\text{P}(\text{R},n \mid \text{NR}) =\pi_{\textrm{R}}(1-\Delta^{n}) ,
\\
\\
\text{P}(\text{R},n \mid \text{R}) =\pi_{\textrm{R}}+\pi_{\textrm{NR}}\Delta^{n} ,
\end{array}
\end{equation}

\noindent where $\Delta:=1-p-q$ and the equilibrium occupancies of the two states are $\pi_{\textrm{R}}=\text{P}(\text{R},n \mid \text{eq})=q/(p+q)$ and $\pi_{\textrm{NR}}=\text{P}(\text{NR},n \mid \text{eq})=p/(p+q)$. The complementary probabilities to be in the non-reactive state are given by

\begin{equation}  \label{eq:100}
\begin{array}{ll}
\text{P}(\text{NR},n \mid \text{NR}) =\pi_{\textrm{NR}}+\pi_{\textrm{R}}\Delta^{n}  ,
\\
\\
\text{P}(\text{NR},n \mid \text{R}) =\pi_{\textrm{NR}}(1-\Delta^{n})  .
\end{array}
\end{equation}

\noindent In case of symmetric internal dynamics, $q=p$, Eqs. (\ref{eq:1}) and (\ref{eq:100}) simplify to

\begin{equation}  \label{eq:2}
\begin{array}{cc}
\text{P}(\text{R},n \mid \text{NR})=\text{P}(\text{NR},n \mid \text{R})  =\frac{1}{2}(1-\Delta^{n}) ,
\\
\\
\text{P}(\text{NR},n \mid \text{NR})=\text{P}(\text{R},n \mid \text{R})=\frac{1}{2}(1+\Delta^{n}),
\end{array}
\end{equation}

\noindent where $\Delta=1-2p$ and $\pi_{\textrm{R}}=\pi_{\textrm{NR}}=\frac{1}{2}$.

\subsection*{The Renewal Approach}

\noindent We now set out to find  the statistics of the discrete random reaction time $N(\Vec{x}_{0})$, which is defined as the number of steps it takes for the particle to react with a target that is placed at the origin. This happens when the particle reaches the origin reactive for the first time. Here, we let $\Vec{r}_{0}$ denote the particle's initial position, and $ \omega_{0}$ its initial internal state, which we jointly denote by $\Vec{x}_{0}=(\Vec{r}_{0}, \omega_{0})$. Setting  $\Vec{0}_{\text{NR}} \equiv(\Vec{0},\text{NR}$), we consider first $N(\Vec{0}_{\textrm{NR}})$: the reaction time of a particle that starts at the origin, albeit in the non-reactive state. We observe that $N(\Vec{0}_{\textrm{NR}})$ can be thought of as a gated first-return time to the origin.

Starting at the origin, the particle can leave and return multiple times, but the overall reaction ends only when the particle arrives at the origin in the reactive state. We can thus write the following renewal equation

\begin{equation} \label{eq:3}
N(\Vec{0}_{\textrm{NR}}) = N_{1} + I_{1}\bigg[N_{2}+I_{2}\Big[N_{3}+I_{3}[...]\Big]\bigg],
\end{equation}

\noindent where $N_{i}$ is the number of steps the particle takes to return to the origin for i-th time, and 
\begin{equation} \label{eq:4}
I_{i}=
\begin{cases}
0, \text{\hspace{1ex} if the particle is reactive at the i-th return,}\\
1, \text{\hspace{1ex} otherwise,}
\end{cases}
\end{equation}
are indicator random variables.  Since the walk is Markovian and time-homogeneous, return processes are independent of each other and statistically identical. Thus,  $\{N_{1},N_{2},...\}$ are independent and identically distributed (IID) copies of a generic return-time which we will henceforth denote $N_{FR}$ (where ``FR" stand for First-Return). Similarly, $\{I_{1},I_{2},...\}$ are IID copies of a generic indicator, which we will henceforth denote $I_{FR}$. To see that the indicators are IID, recall that the internal dynamics are also taken to be Markovian, so that the future internal state depends solely on the current one. Furthermore, note that in the beginning of each return process the overall state of the system is the same: the particle is at the origin in the non-reactive state (otherwise, reaction would have occurred). Indeed, this is exactly the renewal property that allows for the solution presented below.

\begin{figure}[t]
\begin{centering}
\includegraphics[width=1\linewidth]{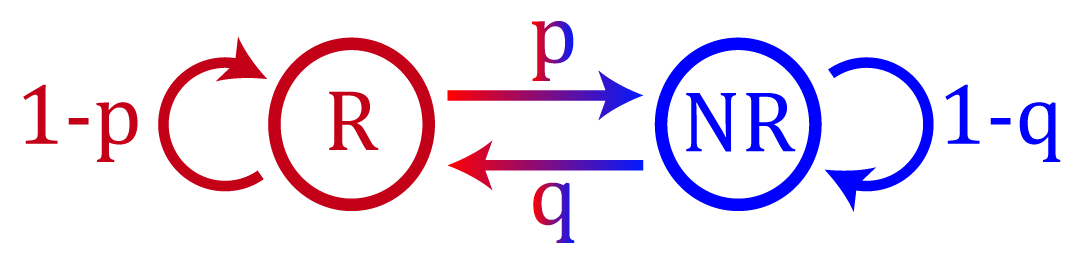}
\caption{Transition probability graph for a discrete-time two-state system. Here, (R) stand for reactive and (NR) stands for non-reactive (NR).}
\end{centering}
\end{figure} 

\subsection*{ Distribution of the Reaction Time}
\noindent In principle, Eq. (\ref{eq:3}) allows us to compute the distribution of $N(\Vec{0}_{\textrm{NR}})$ directly. For example, the probability that $N(\Vec{0}_{\textrm{NR}})=4$ can be found by summing over all possible combinations with this reaction time: $N_1=4$, $I_1=0$ being the first; and $N_1=2$, $I_1=1$ followed by $N_2=2$, $I_2=0$ being the second. However, this straightforward approach complicates rather quickly, yielding cumbersome expressions. Instead, we use a trick of the trade and compute the Z-transform, a.k.a the generating function, of $N(\Vec{0}_{\textrm{NR}})$:

\begin{equation} \label{eq:5}
\tilde{N}(\Vec{0}_{\textrm{NR}},z)=\braket{z^{N(\Vec{0}_{\textrm{NR}})}}
=\braket{z^{N_{1}} z^{I_{1}\Big[N_{2}+I_{2}\big[...\big]\Big]}}.
\end{equation}

\noindent By using the law of iterated expectations:

\begin{equation} \label{eq:6}
\tilde{N}(\Vec{0}_{\textrm{NR}},z)=\braket{\braket{
z^{N_{1}} z^{I_{1}\Big[N_{2}+I_{2}\big[...\big]\Big]} \mid N_{1} = n
}}_{N_{1}}     
\end{equation}

\noindent For a given value of $N_{1}$, the inner expectation in Eq. (\ref{eq:6}) reads 

\begin{equation} \label{eq:7}
\begin{array}{ll}
z^{n} 
\braket{z^{I_{1}\Big[N_{2}+I_{2}\big[...\big]\Big]} \mid N_{1} = n}  
      \\ \\
 = z^{n} \text{P}(\text{R},n \mid \text{NR}) + z^{n} \text{P}(\text{NR},n \mid \text{NR}) \braket{z^{N_{2}+I_{2}\big[...\big]} },
\end{array}
\end{equation}

\noindent where we have used the fact that $I_{1}=0$ with probability $\text{P}(\text{R},n \mid \text{NR})$ and $I_{1}=1$ with probability $\text{P}(\text{NR},n \mid \text{NR})$. Plugging this result back into Eq. (\ref{eq:6}), we obtain

\begin{equation}  \label{eq:8}
\begin{array}{cc}
  \tilde{N}(\Vec{0}_{\textrm{NR}},z)=
 \braket{z^{N_{1}}\text{P}(\text{R},N_{1} \mid \text{NR})}     
       \\ \\
 +\braket{z^{N_{1}}\text{P}(\text{NR},N_{1} \mid \text{NR})} \braket{
z^{N_{2}+I_{2}[...]}}.       
\end{array}
\end{equation}

We now observe that due to statistical identity of the return processes $\braket{z^{N_{2}+I_{2}[...]}}=\braket{z^{N(\Vec{0}_{\textrm{NR}})}}=\tilde{N}(\Vec{0}_{\textrm{NR}},z)$. Then, by plugging the relations in Eqs. (\ref{eq:1}) and (\ref{eq:100}) into Eq. (\ref{eq:8}) and rearranging we get

\begin{equation} \label{eq:9}
\tilde{N}(\Vec{0}_{\textrm{NR}},z)=
\frac{\tilde{N}_{\textrm{FR}}(z) -\tilde{N}_{\textrm{FR}}(\Delta z)} {1+K_{eq}\Big[1-\tilde{N}_{\textrm{FR}}(z)\Big]-\tilde{N}_{\textrm{FR}}(\Delta z)},
\end{equation}

\noindent where $K_{eq}=p/q$ and $\tilde{N}_{\textrm{FR}}(z)$ is the Z-transform of $N_{FR}$, which was defined below Eq. (\ref{eq:4}) as the time it takes the particle to return to the origin. This is the first main result of this manuscript, as it shows that the gated reaction time when starting at the origin at the non-reactive state can be expressed in terms of the corresponding ungated first-return time. Thus, via Eq. (\ref{eq:9}), the solution of a gated problem is reduced to that of the much simpler ungated problem. We stress that this result is general and holds for any gated reaction in discrete time and space. 

In the case of symmetric internal dynamics, $K_{eq}=1$, Eq. (\ref{eq:9}) simplifies to

\begin{equation} \label{eq:10}
\tilde{N}(\Vec{0}_{\textrm{NR}},z)=
\frac{\tilde{N}_{\textrm{FR}}(z) -\tilde{N}_{\textrm{FR}}(\Delta z)} {2-\tilde{N}_{\textrm{FR}}(z)-\tilde{N}_{\textrm{FR}}(\Delta z)},
\end{equation}

\noindent where $\Delta = 1-2p$. There is an interesting case in which Eq. (\ref{eq:10}) displays a striking symmetry under the transformation $p \mapsto 1-p$, which is here equivalent to $\Delta \mapsto -\Delta$. Consider a situation where return to the origin can only be made in an even number of steps, as happens for example in a simple random walk in $1D$. In this case $\tilde{N}_{\textrm{FR}}(z)$ is an even function since all terms with odd powers of $z$ vanish in the Z-transform. We thus have $\tilde{N}_{\textrm{FR}}(\Delta z)=\tilde{N}_{\textrm{FR}}(-\Delta z)$, which means  that $\tilde{N}(\Vec{0}_{\textrm{NR}},z)$ is itself invariant under the transformation $p \mapsto 1-p$. In section IV, we return to this symmetry and its consequences when we treat the problem of a 1D random walk with a gated target.  

Now that we have an expression for $\tilde{N}(\Vec{0}_{\textrm{NR}},z)$, we can generalize to an arbitrary  initial condition $\Vec{x}_{0}=(\Vec{r}_{0}, \omega_{0})$, where we recall that $\Vec{r}_{0}$ denotes the particle's initial position, and $ \omega_{0}$ its initial internal state. To compute the reaction time, we observe that a particle which start at $\Vec{r_{0}}$ must first arrive to the origin. Letting $N_{FP}(\Vec{r_{0}})$ denote the first-passage time from $\Vec{r_{0}}$ to the origin, we now consider what happens next. When arriving at the origin there are two options: i) The particle is in the reactive state and a reaction takes place immediately; ii) The particle is the non-reactive state and the reaction does not take place. The second scenario brings us back to the situation described and solved above: an additional time $N(\Vec{0}_{\textrm{NR}})$ is required for the reaction to complete. The total reaction time is thus 
\begin{equation} \label{eq:11}
N(\Vec{x}_{0}) = N_{FP}(\Vec{r}_{0}) + I_{FP}N(\Vec{0}_{\textrm{NR}}),
\end{equation}

\noindent where $I_{FP}$ is defined as in Eq. (\ref{eq:4}), but for first-passage rather than return.

We proceed as before, Z-transforming Eq. (\ref{eq:11}) we obtain

\begin{equation} \label{eq:12}
\begin{array}{cc}
\tilde{N}(\Vec{x}_{0},z)=
\braket{z^{N_{FP}(\Vec{r}_{0})}\text{P}(\text{R},N_{FP}(\Vec{r}_{0}) \mid  \omega_{0})}      \\ \\
+\braket{z^{N_{FP}(\Vec{r}_{0})}\text{P}(\text{NR},N_{FP}(\Vec{r}_{0}) \mid  \omega_{0})} \tilde{N}(\Vec{0}_{\textrm{NR}},z),   
\end{array}
\end{equation}

\noindent and by plugging the relevant expressions in Eqs. (\ref{eq:1}) and (\ref{eq:100}) into Eq. (\ref{eq:12}) we get:
\begin{equation} \label{eq:13}
 \begin{array}{cc}
 \tilde{N}(\Vec{x}_{0},z)= \tilde{N}_{FP}(\Vec{r}_{0},z)\Big[\pi_{R} 
 +\pi_{NR}\tilde{N}(\Vec{0}_{\textrm{NR}},z)\Big]
 \\ \\
 + I_{\omega_{0}}(1-\pi_{ \omega_{0}}) \tilde{N}_{FP}(\Vec{r}_{0},\Delta z) \Big[ \tilde{N}(\Vec{0}_{\textrm{NR}},z) - 1 \Big],
  \end{array}
\end{equation}
\noindent where $\tilde{N}_{FP}(\Vec{r_0},z)$ and $\tilde{N}_{FP}(\Vec{r_0},\Delta z)$ are the Z-transforms of $N_{FP}(\Vec{r}_{0})$ evaluated at $z$ and $\Delta z$ respectively, $\tilde{N}(\Vec{0}_{\textrm{NR}},z)$ was defined in Eq. (\ref{eq:9}), and where 

\begin{equation} \label{eq:115}
 I_{\omega_{0}} = \begin{cases}
 1,  \hspace{17pt} \omega_{0}=\text{NR},\\
 0, \hspace{17pt}  \omega_{0}=\text{eq},\\
 -1, \hspace{10pt}  \omega_{0}=\text{R}.
 \end{cases}    
\end{equation}

Equation (\ref{eq:13}) is the second main result of this manuscript. It shows that the gated reaction time can always be expressed in terms of the corresponding, and much simpler, ungated first-return and passage times. Note that for an equilibrium initial condition of the internal state, $\omega_{0}=\text{eq}$, the result simplifies considerably as the second row of Eq. (\ref{eq:13}) vanishes. Also, when the starting position  $\Vec{r}_{0}$ is far away from the origin, the internal state equilibrates before the particle arrives there and all information on the internal initial condition is lost. Mathematically, this can be seen by observing that ${N}_{FP}(\Vec{r}_{0})$ is then typically very large, making $\tilde{N}_{FP}(\Vec{r}_{0},\Delta z)=\Braket{(\Delta z)^{{N}_{FP}(\Vec{r_0})}}$ negligible with respect to $\tilde{N}_{FP}(\Vec{r}_{0}, z)=\Braket{z^{{N}_{FP}(\Vec{r_0})}}$. Thus, this case if effectively equivalent to that of the equilibrium initial condition.

\subsection*{Mean and Variance}

\noindent  The Z-transform is a moment generating function, so the moments of $N(\Vec{0}_{\textrm{NR}})$ can be calculated directly from Eq. (\ref{eq:9}) by taking derivatives.\cite{klafter2011first} Nonetheless, we think the following independent calculation of the mean of $N(\Vec{0}_{\textrm{NR}})$ is instructive. To find $\braket{N(\Vec{0}_{\textrm{NR}})}$, we take the mean of both sides of Eq. (\ref{eq:3})

\begin{equation} \label{eq:14}
\braket{N(\Vec{0}_{\textrm{NR}})} = \braket{N_{1}} + \braket{I_{1}N_{2}} + \braket{I_{1}I_{2}N_{3}} + \dots
\end{equation}

\noindent Using the independence of the random variables at hand we get

\begin{equation} \label{eq:15}
\braket{N(\Vec{0}_{\textrm{NR}})} = \braket{N_{1}} + \braket{I_{1}}\braket{N_{2}} + \braket{I_{1}}\braket{I_{2}}\braket{N_{3}} + \dots
\end{equation}

\noindent Now recalling that the return times are IID, we have $\braket{N_{i}}=\braket{N_{FR}}$ and $\braket{I_{i}}=\braket{I_{FR}}$, and thus

\begin{equation} \label{eq:16}
\begin{array}{ll}
   \braket{N(\Vec{0}_{\textrm{NR}})} = \braket{N_{FR}} \sum_{k=0}^{\infty}\braket{I_{FR}}^{k}   
   \\ \\
= \braket{N_{FR}}\sum_{k=0}^{\infty}  \braket{\text{P}(\text{NR},N_{FR} \mid \text{NR})}^{k} 
 = \frac{\braket{N_{FR}}}{1-\braket{\text{P}(\text{NR},N_{FR} \mid \text{NR})}}.
\end{array}
\end{equation}

\noindent Observe that the probability that the particle returns to the origin in the reactive state is given by  $\braket{\text{P}(\text{R},N_{FR} \mid \text{NR})}=1-\braket{\text{P}(\text{NR},N_{FR} \mid \text{NR})}$. Thus, on average, the particle returns to the origin $\braket{\text{P}(\text{R},N_{FR} \mid \text{NR})}^{-1}$ times, with each return taking $\braket{N_{FR}}$ time units on average, before a reaction occurs. Recalling Eq. (\ref{eq:1}), we have $\Braket{ \text{P}(\text{R},N_{FR} \mid \text{NR})} = \pi_{\textrm{R}}[1-\tilde{N}_{\textrm{FR}}(\Delta)]$. Plugging this result into Eq. (\ref{eq:16}) and rearranging we conclude that

\begin{equation} \label{eq:18}
\braket{N(\Vec{0}_{\textrm{NR}})} = \frac{\pi^{-1}_{\textrm{R}} }{1- \tilde{N}_{\textrm{FR}}(\Delta)}\braket{N_{FR}}.
\end{equation}

\noindent Once again, note that this result is completely general.

In the same manner one can average Eq. (\ref{eq:11}) to get

\begin{equation} \label{eq:19}
\braket{N(\Vec{x}_{0})}=\braket{N_{FP}(\Vec{r}_{0})} + \braket{I_{FP}}\braket{N(\Vec{0}_{\textrm{NR}})}.
\end{equation}

\noindent Noting that $\Braket{I_{FP}} =\Braket{\text{P}(\text{NR},{N_{FP}(\Vec{r}_{0})} \mid  \omega_{0})}$, we can plug in the relevant term from Eqs. (\ref{eq:1}) and (\ref{eq:100}) for $ \omega_{0} \in \{\text{R, NR, eq}\} $ to calculate the corresponding $\Braket{I_{FP}}$ and plug it into Eq. (\ref{eq:19}). This gives

\begin{equation} \label{eq:21}
\begin{array}{ll}
\braket{N(\Vec{x}_{0})}=\braket{N_{FP}(\Vec{r}_{0})} +  \pi_{\textrm{NR}} \braket{N(\Vec{0}_{\textrm{NR}})}     \\ \\
+ I_{\omega_{0}} (1-\pi_{ \omega_{0}})\tilde{N}_{FP}(\Vec{r}_{0},\Delta)\braket{N(\Vec{0}_{\textrm{NR}})},      
\end{array}
\end{equation}

\noindent where $I_{\omega_{0}}$ was defined in Eq. (\ref{eq:115}), and where $\tilde{N}_{FP}(\Vec{r_0},\Delta)=\Braket{\Delta^{{N}_{FP}(\Vec{r_0})}}$ is the Z-transform of $N_{FP}(\Vec{r}_{0})$ evaluated at $\Delta$. Here as well, if the particle starts far enough such that the internal state equilibrates before the particle arrives at the origin, the Z-transform is negligible and the mean reaction time in Eq. (\ref{eq:21}) becomes independent of the initial internal state (Clearly $\Delta^{N_{FP}(\Vec{r}_{0})}$ is then typically very small, as $\abs{\Delta} < 1$ for all cases but the redundant cases of $p=q=0$ or $p=q=1$). The result is then equivalent to that obtained in the case of $\omega_{0}=\text{eq}$, and is very intuitive: The first arrival to the origin occurs after $\braket{N_{FP}(\Vec{r}_{0})}$ on average. With probability $\pi_{R}$ the first arrival is fertile, so it is also the last one. With the complementary probability, $\pi_{NR}$, the first arrival is infertile, and the reaction continues for an additional $\braket{N(\Vec{0}_{\textrm{NR}})}$ steps on average. 

As mentioned in the beginning of this section, we could have reached Eq. (\ref{eq:18}) by using the fact that the Z-transform in Eq. (\ref{eq:9}) is in fact a moment generating function. For the mean one employs $\braket{N(\Vec{0}_{\textrm{NR}})} = \frac{d\tilde{N}(\Vec{0}_{\textrm{NR}},z)}{d z} \Big|_{z = 1}$, which reproduces Eq. (\ref{eq:9}) with little algebra. For the second moment, we have $\braket{N^{2}(\Vec{0}_{\textrm{NR}})} = \frac{d^2\tilde{N}(\Vec{0}_{\textrm{NR}},z)}{d z^2} \Big|_{z = 1}+ \braket{N(\Vec{0}_{\textrm{NR}})}$. Recalling $\text{Var}(N(\Vec{0}_{\textrm{NR}}))=\braket{N^{2}(\Vec{0}_{\textrm{NR}})} - \braket{N(\Vec{0}_{\textrm{NR}})}^{2}$, we obtain

\begin{equation} \label{eq:22}
\begin{array}{ll}
\text{Var}(N(\Vec{0}_{\textrm{NR}})) =  \frac{\pi^{-1}_{\textrm{R}}}{1- \tilde{N}_{\textrm{FR}}(\Delta)} \Big[ \braket{N^{2}_{FR}} +
\\ \\
 \frac{K_{eq} - 1}{1- \tilde{N}_{\textrm{FR}}(\Delta)}\braket{N_{FR}}^2+\frac{\Delta }{1- \tilde{N}_{\textrm{FR}}(\Delta)} \frac{d\tilde{N}_{\textrm{FR}}(z)}{dz} \Big|_{z=\Delta}\Big],      
\end{array}
\end{equation}

\noindent and a result for $\text{Var}(N(\Vec{x}_{0}))$ can be obtained in a similar way.
\subsection*{When the Mean Diverges}

\noindent When deriving Eq. (\ref{eq:18}) we implicitly assumed that $\braket{N_{FR}}$ is finite. The expression is of course meaningless if $\braket{N_{FR}}$ diverges as in this case $\braket{N(\Vec{0}_{\textrm{NR}})}$ must also diverge. This is easy to see from Eq. (\ref{eq:3}) which implies that the gated reaction time $N(\Vec{0}_{\textrm{NR}})$ is always larger or equal to the return time $N_{FR}$. Nonetheless, even when the mean diverges, Eq. (\ref{eq:9}) which gives the Z-transform of $N(\Vec{0}_{\textrm{NR}})$ is still valid and of use. This is not to say that the inverse transformation is a simple task, or that the distribution can always be obtained in closed form. Rather, we hereby focus on the long-time asymptotics, and show that when the underlying ungated process is controlled by a heavy power-law tail, this tail is 
inherited by the corresponding gated process. More precisely, we show that the power law remains the same and that the asymptotics differ only in the corresponding prefactor, which is determined exactly. 

Consider the ungated process, and let $f_{0}(n)$ be the PMF of the random variable $N_{FR}$, i.e., $f_{0}(n)$ is the probability that the first return to the origin occurs exactly after n steps. The survival is defined to be $S(n)=\sum_{n+1}^{\infty}f_{0}(n)$, namely the probability that the particle did \textit{not} return to the origin by step $n$. The following relation between the Z-transforms $\tilde{S}(z)=\sum_{n=0}^{\infty} S(n) z^{n}$ and $\tilde{N}_{\textrm{FR}}(z)=\sum_{n=0}^{\infty} f_0(n) z^{n}$ holds \cite{bonomo2021first}

\begin{equation}  \label{eq:23}
    \tilde{S}(z) = \frac{1-\tilde{N}_{\textrm{FR}}(z)}{1-z}.
\end{equation}

\noindent The derivation of this relation can found in Appendix C of Ref. 37.

Let us now assume that $f_{0}(n)$ is monotonic (at least asymptotically) with a heavy power-law tail such that for $n \gg 1$ we have
 
\begin{equation}  \label{eq:24}
    f_{0}(n)  \simeq C n^{-\gamma},
\end{equation}

\noindent with $1<\gamma<2$; and note that in this case  $\braket{N_{FR}}$ diverges. The asymptotics of the survival function is then given by  $S(n) \simeq \frac{C}{\gamma-1} n^{1-\gamma}$. By applying the Tauberian theorem,\cite{klafter2011first} we obtain $\tilde{S}(z)  \simeq \frac{C \Gamma (2-\gamma)}{\gamma-1} (1-z)^{\gamma-2}$ in the limit $z \to 1$, where $\Gamma(\cdot)$ is the Gamma function. Using Eq. (\ref{eq:23}), we obtain

\begin{equation}  \label{eq:25}
    \tilde{N}_{\textrm{FR}}(z) = 1 - \frac{C \Gamma (2-\gamma)}{\gamma-1} (1-z)^{\gamma-1}.
\end{equation}

\noindent To emphasize, we have found that if the long-time asymptotics of the first return time is given by Eq. (\ref{eq:24}), then its Z-transform is given by Eq. (\ref{eq:25}). 

Plugging the result of Eq. (\ref{eq:25}) into Eq. (\ref{eq:9}), and taking the limit $z \to 1$, we obtain (after some algebra)

\begin{equation}  \label{eq:26}
    \tilde{N}(\Vec{0}_{\textrm{NR}},z) \simeq 1 - \frac{C\Gamma(2-\gamma)}{\gamma-1}\frac{\pi^{-1}_{\textrm{R}}}{1-\tilde{N}_{FR}(\Delta)}(1-z)^{\gamma-1}.
\end{equation}

\noindent Equation (\ref{eq:26}) is of the same form of Eq. (\ref{eq:25}), up to a different pre-factor. Thus, their inversions are also the same, up to the same pre-factor:

\begin{equation}  \label{eq:27}
      f_{0,\text{NR}}(n) \simeq \frac{ \pi^{-1}_{\textrm{R}}}{1-\tilde{N}_{FR}(\Delta)} C n^{-\gamma},
\end{equation}

\noindent where $f_{0,\text{NR}}(n)$ is the PMF of the gated reaction time $N(\Vec{0}_{\textrm{NR}})$. We have thus proved that the asymptotic power law of the ungated problem is inherited by its gated counterpart. The pre-factor is changed, however, by a factor of $\frac{\pi^{-1}_{\textrm{R}}}{1-\tilde{N}_{FR}(\Delta)}$ as can be seen by comparing Eqs. (\ref{eq:24}) and (\ref{eq:27}). Note that this is the same factor appearing in Eq. (\ref{eq:18}).

There is one last subtle point to note here. In Eq. (\ref{eq:24}) we have assumed that $f_{0}(n)$ is  asymptotically monotonic. In reality, discrete time random walks on networks often possess non-monotonic first-return PMFs. For example, in the simple 1D random walk return on odd number of steps has zero probability. However, the even sequence $\{f_{0}(2n)\}$ is indeed monotonic. In this case, Eq. (\ref{eq:24}) should be understood as an average of subsequent even and odd steps. Then, $f_{0}(n)$ is zero for odd $n$ and two times the quantity of Eq. (\ref{eq:24}) for even $n$. Alternatively, one can solve directly for the sequence $\{f_{0}(2n)\}$, and obtain the same factor of 2. In more complicated networks, one should be careful and apply similar considerations if needed.

\subsection*{Transient Power-Law Behaviour in High Crypticity}
\noindent Let us continue with the exact same scenario of the heavy-tailed $f_{0}(n)$ given in Eq. (\ref{eq:24}). But now, let us also assume high crypticity, i.e., $K_{eq} = p/q \gg 1$. To study the transition to the long-time asymptotics under the high-cripticity assumption, we take $z$ close to $1$ in Eq. (\ref{eq:9}), but keep $K_{eq}$ large enough such that  $K_{eq}\Big[1-\tilde{N}_{\textrm{FR}}(z)\Big] \gg 1$ holds. Using $\tilde{N}_{\textrm{FR}}(z)$ from Eq. (\ref{eq:25}) we obtain 

\begin{equation}  \label{eq:50}
 \tilde{N}(\Vec{0}_{\textrm{NR}},z) \simeq  \frac{1-\tilde{N}_{FR}(\Delta)} {K_{eq} \frac{C \Gamma(2-\gamma)}{\gamma-1}}\frac{1}{(1-z)^{\gamma-1}}.
\end{equation}

\noindent Applying the Tauberian theorem gives

\begin{equation}  \label{eq:51}
  f_{0,\text{NR}}(n) \simeq  \frac{\Big[1-\tilde{N}_{FR}(\Delta)\Big](\gamma-1)} {K_{eq} C \Gamma(2-\gamma) \Gamma(\gamma-1)} n^{\gamma-2}.
\end{equation}

Equation (\ref{eq:51}) means that under high crypticity there is a transient regime, possibly very long, which is governed by a $\sim n^{\gamma-2}$ power law, before the terminal $\sim n^{-\gamma}$ asymptotics enters. It is interesting to note that by using Euler's reflection formula, Eq. (\ref{eq:51}) can be re-written as  

\begin{equation}  \label{eq:52}
  f_{0,\text{NR}}(n) \simeq  \frac{\Big[1-\tilde{N}_{FR}(\Delta)\Big]} {K_{eq} C} \frac{x\text{sin}(\pi x)}{\pi} n^{x-1},
\end{equation}

\noindent where we have denoted $x:=\gamma-1$, such that $0<x<1$. We can thus easily see that the pre-factor is maximized for the case $x=1/2$ (which corresponds to $\gamma=3/2$), and that it decreases towards zero for both larger and smaller values of $x$, but in an asymmetric manner, as the sine function is multiplied by $x$.

Finally, note that by employing the same technique, one can obtain the analogous result for the continuous-time case, starting from Eq. 7 of Ref. 34.


\section*{III. Example: Gated Reaction with Unidirectional Motion on a Ring}

\noindent To illustrate the framework developed above, consider the following problem: A particle is bound to a ring lattice of $N$ sites. The particle's motion is unidirectional, namely at every time step, the particle takes one step clockwise. In addition, the particle switches between a reactive (red) and non-reactive (blue) states as described in Fig. 3. Starting the particle at the origin, in the non-reactive state, we are interested in the gated reaction time. Namely, the time it takes the particle to return to the origin in the reactive state.  

Recall that in the previous section we showed that the  statistics of the gated reaction time $N(\Vec{0}_{\textrm{NR}})$ can be expressed in terms of the corresponding ungated reaction time $N_{FR}$, i.e., the first return time to the origin. Specifically, in the case considered above, calculation of the ungated reaction time is extremely simple. Since the particle's motion is unidirectional, it will return to the origin after taking exactly N steps. In other words, we have $\braket{N_{FR}}=N$ and $\text{Var}(N_{FR})=0$, which means $\tilde{N}_{\textrm{FR}}(\Delta) = \Delta^N$. The mean of the gated reaction time, which is less trivial, now follows directly from Eq. (\ref{eq:18})  

\begin{figure}[t]
\begin{centering}
\includegraphics[width=0.8\linewidth]{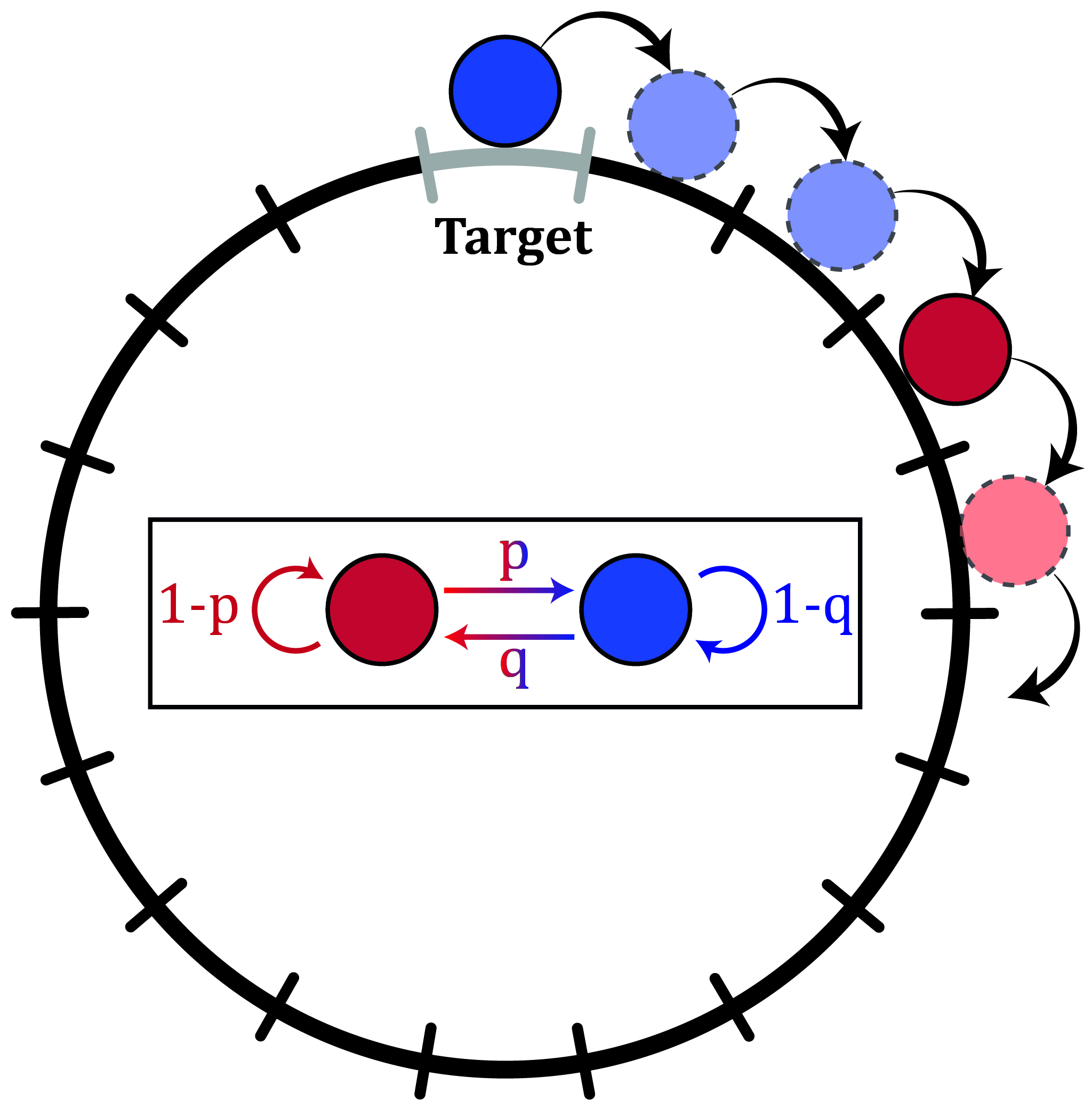}
\caption{An illustration of a gated reaction on a ring. A particle transitions between a reactive state (red) and a non-reactive state (blue) according to the Markov chain illustrated inside the square box. In addition, at every time step, the particle jumps to its right neighbouring site. The particle starts on the target in the non-reactive state. A reaction occurs when the particle first returns to the target in the reactive state.}
\end{centering}
\end{figure}

\begin{equation} \label{eq:28}
\braket{N(\Vec{0}_{\textrm{NR}})} = \frac{\pi^{-1}_{\textrm{R}} }{1- \Delta^{N}} N,
\end{equation}

\noindent which for the symmetric case, $p=q$, boils down to 

\begin{equation} \label{eq:29}
\braket{N(\Vec{0}_{\textrm{NR}})} = \frac{2}{1- (1-2p)^{N}} N.
\end{equation}

A few interesting features can be deduced from Eq. (\ref{eq:28}). For $-1 < \Delta < 1$, we have $\braket{N(\Vec{0}_{\textrm{NR}})} \to \pi^{-1}_{\textrm{R}} N$ as $N \to \infty$. To understand this, observe that when $N$ is large enough the internal state has enough time to  equilibrate within a single revolution of the ring. Thus, the probability to find the particle in the reactive state when it returns to the origin is the equilibrium probability $\pi_{\textrm{R}}$. For example, when $p=q$ we have $\pi_{\textrm{R}}=\frac{1}{2}$ and so $\braket{N(\Vec{0}_{\textrm{NR}})} \to 2 N$. Thus, on average, the particle misses the target in half of its returns, yielding a mean reaction time that is twice that of the ungated problem. Note that this limit is independent on the value of $\Delta$, and that it can give a fair approximation for $\braket{N(\Vec{0}_{\textrm{NR}})}$ even for moderate values of $N$. This is illustrated in Fig. 4, where we compare numerical simulations with theory for the symmetric, $p=q$, case. Observe that $\braket{N(\Vec{0}_{\textrm{NR}})}\simeq2N$, except for extreme values of $p$, where this approximation is expected to break. 

The mean gated reaction time displays clear parity effects with regard to the size of the ring. This happens because the internal gating dynamics can be in, or out of, phase with the  molecular encounters of the particle with the target, thus leading to resonance and anti-resonance effects. For example, in the symmetric case illustrated in Fig. 4, two distinct behaviours arise in the limit $p \to 1$: i) If $N$ is even the mean reaction time diverges. Indeed, for $p=1$ the particle changes its internal state with every step, and given that it started non-reactive it is sure to return to the origin non-reactive. ii) Contrary, when $N$ is odd, the mean reaction time approaches $N$ because the particle is almost sure to transition to the reactive state as it steps back into the origin at the end of a revolution. The same effect is apparent in Fig. 5a and 5b, where either $q$ or $p$ is set constant to a high value of 0.9 and the other parameter is swept across the unit interval. As this parameter approaches 1, we reach a scenario in which both $p$ and $q$ are large and similar parity effects are observed.

\begin{figure}[t]
\begin{centering}
\includegraphics[width=0.9\linewidth]{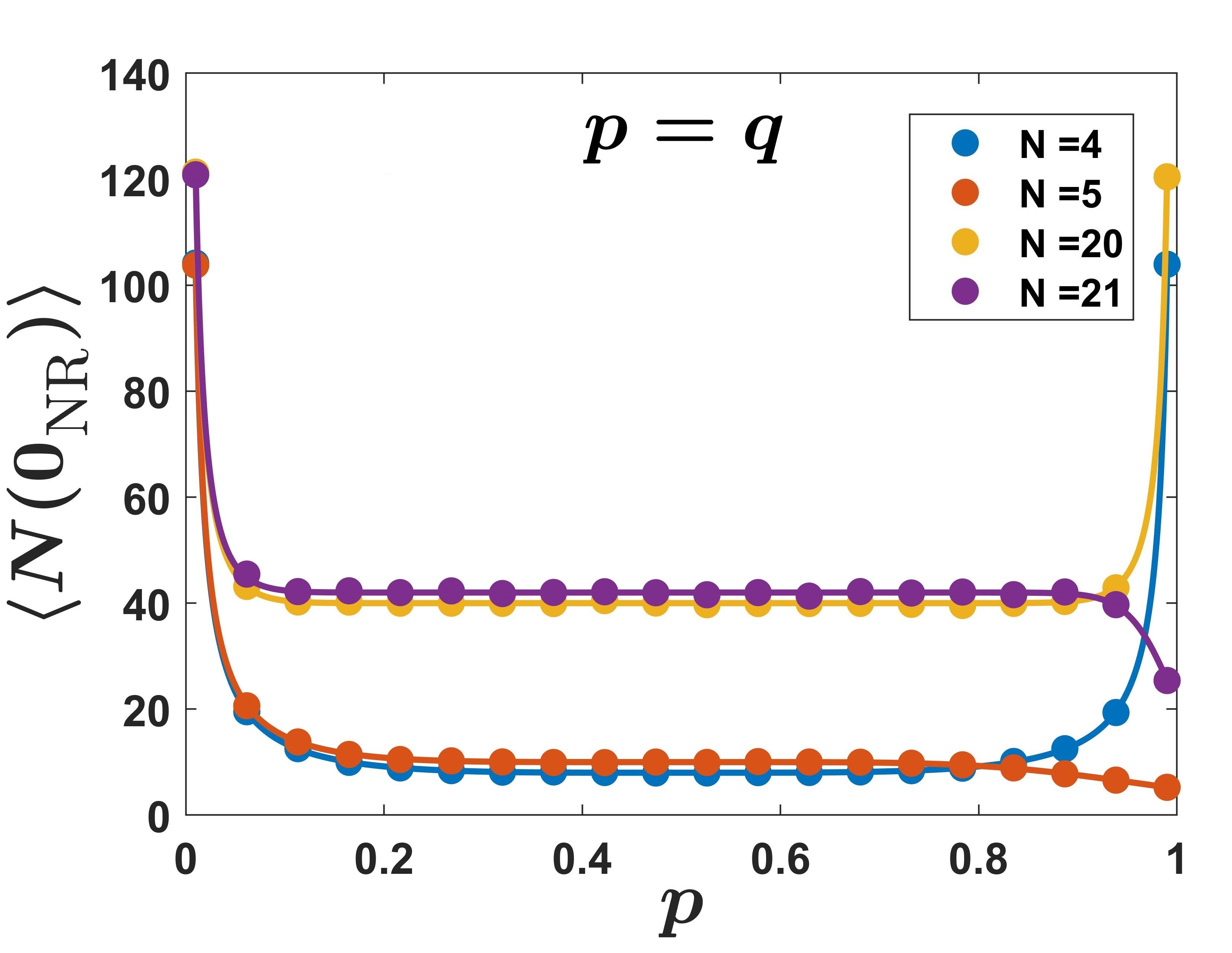}
\caption{The mean gated reaction time $\braket{N(\Vec{0}_{\textrm{NR}})}$ vs. the transition probability $p$ for the reaction in Fig. 3. Here, we consider symmetric internal dynamics ($p=q$) and various lattice sizes $N=$ 4, 5, 20 and 21. Solid lines  represent the theory coming from Eq. (\ref{eq:29}), and full circles come from numeric simulations. Observe that $\braket{N(\Vec{0}_{\textrm{NR}})}\simeq2N$, except for extreme values of $p$, where this approximation is expected to break. As $p \to 0$, the probability to leave the initial non-reactive state vanishes and the mean reaction time diverges. In the other limit, $p \to 1$, clear parity effects arise: for even $N$ the mean diverges, while for odd $N$ the mean is approximately $N$ itself.}
\end{centering}
\end{figure}

\begin{figure*}
\begin{centering}
\includegraphics[width=0.9\linewidth]{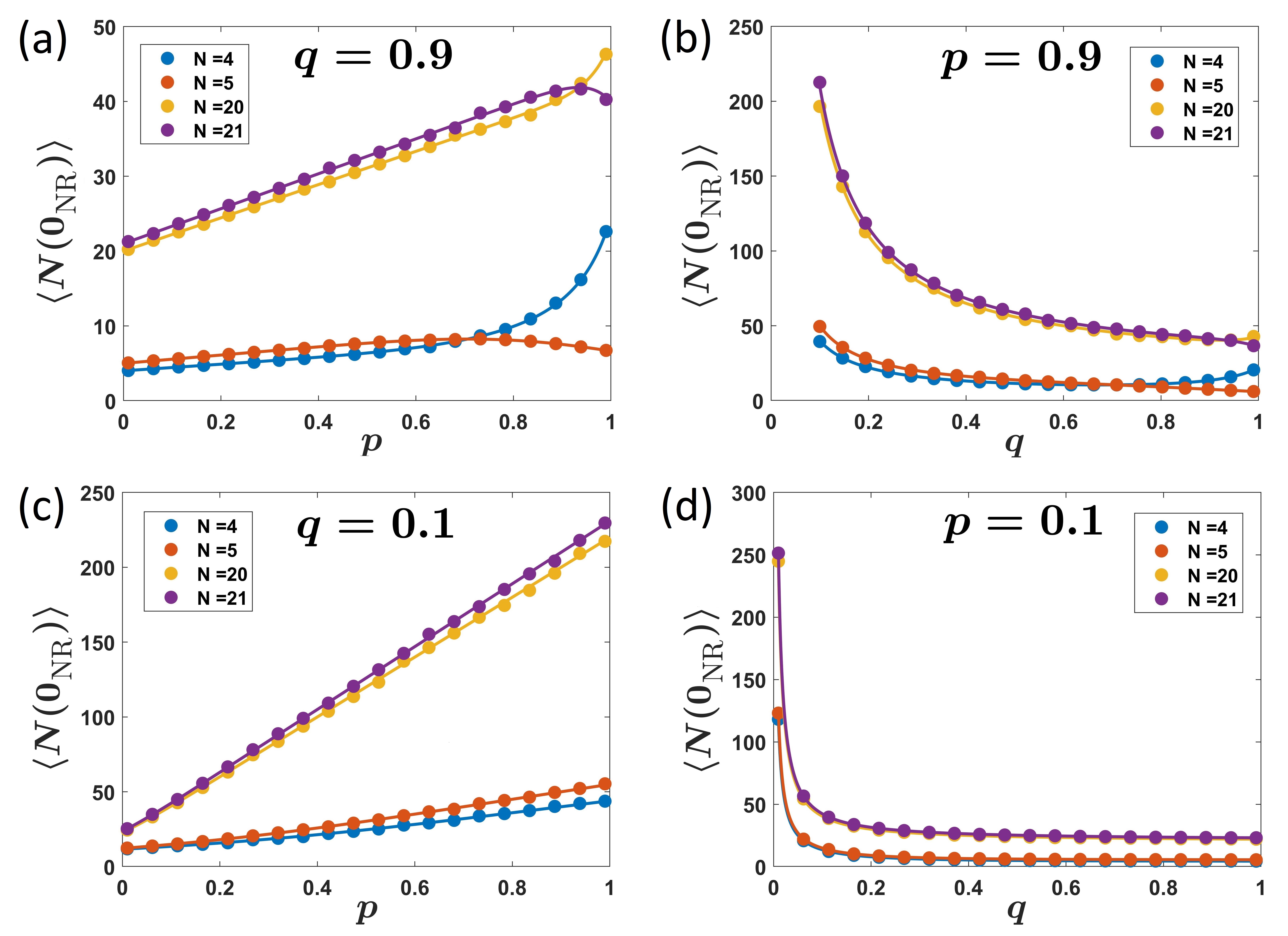}
\caption{The mean gated reaction time $\braket{N(\Vec{0}_{\textrm{NR}})}$ vs. the transition probability $p$ or $q$ for the reaction in Fig. 3, for various lattice sizes $N=$ 4, 5, 20 and 21. Solid lines  represent the theory coming from Eq. (\ref{eq:28}), and full circles come from numeric simulations. In (a) we set $q=0.9$ and vary $p$, in (b) we set $p=0.9$ and vary $q$, in (c) we set $p=0.1$ and vary $p$, and in (d) we set $q=0.1$ and vary $q$.}
\end{centering}
\end{figure*}

An interesting scenario is that of Fig. 5c, where $q$ is set to the low value of 0.1. If $p$ is also low, once the particle becomes reactive it will stay that way for a long time. We thus observe a mean reaction time which corresponds to the number of revolutions needed for the particle to become reactive for the first time. However, as $p$ increases it becomes much more likely that the particle reverts back to the non-reactive state before completing a revolution. As a result, the mean reaction time is greatly prolonged. In Fig. 5a we set $q$ to high value of 0.9 and the effects are much less pronounced: When $p$ is low, the mean reaction time tends to the size of the ring, as it is very likely that the particle will become reactive within the first revolution and stay that way. As $p$ is increased, we again have the same effect of prolonged reaction times, but to a lesser extent (note differences in slopes compared to Fig. 5c).

Lastly, in Fig. 5d we set $p=0.1$ and vary $q$. When $q \to 0$ the probability to leave the initial non-reactive state vanishes and the mean reaction time diverges. When $q \to 1$ the particle transitions from the non-reactive state to the reactive state very fast relative to opposite transition, and it is very likely to arrive reactive at the end of the first revolution.

We can also obtain the PMF of the reaction time when starting on the target in the non-reactive state. By plugging $\tilde{N}_{\textrm{FR}}(z) = z^N$ in Eq. (\ref{eq:9}) we obtain

\begin{equation}   \label{eq:30}
\tilde{N}(\Vec{0}_{\textrm{NR}},z)=
\frac{z^N - (\Delta z)^N} {1+K_{eq}\Big[1-z^N\Big]-(\Delta z)^N}.
\end{equation}

\noindent By denoting $\zeta=z^{N}$ and recalling Eq. (\ref{eq:1}):  $\text{P}(\text{R},n \mid \text{NR}) =\pi_{\textrm{R}}(1-\Delta^{n})$ we can rewrite Eq. (\ref{eq:30}) as

\begin{equation}   \label{eq:31}
\tilde{N}(\Vec{0}_{\textrm{NR}},\zeta)=
\frac{\text{P}(\text{R},N \mid \text{NR}) \zeta }{1 - \big[1-\text{P}(\text{R},N \mid \text{NR})\big]\zeta},
\end{equation}

\noindent which is a Z-transform of a geometric distribution evaluated at $\zeta$. Here the success probability is $\text{P}(\text{R},N \mid \text{NR})$, as each geometric trial is equivalent to a full revolution of $N$ steps. To understand this result we observe that the particle returns to the target every $N$ steps. The probability of the particle to be reactive on return is exactly $\text{P}(\text{R},N \mid \text{NR})$. If the particle is reactive on return then the walk is over. However if it is non-reactive, the process is renewed and again the probability to be reactive in the next revolution is $\text{P}(\text{R},N \mid \text{NR})$. This is indeed a geometric distribution.

In fact, because the $N_{FR}$ is deterministic the gated problem is effectively reduced to that of a radiating boundary condition, i.e., following each revolution there is a constant probability of reacting. We stress that this reduction is not due to equilibration of the internal state, but rather due to the deterministic return time. This, of course, makes the gated problem considerably easier to solve, but our formalism also allows us to treat cases where the return time fluctuates. For such cases, the reduction above is impossible (unless internal state equilibrium is certain to be reached between any two consecutive returns to the target). In the next section, we consider a problem in which not only that the return time fluctuates --- it varies so greatly that the mean return time diverges.

\section*{IV. Example: Symmetric Random-Walk on a Semi-Infinite Interval with a Gated Boundary}

\noindent Consider a random walk on a semi-infinite one-dimensional lattice with a two-state gated boundary as depicted in Fig. 6.  Assuming the particle starts at an initial location $x_{0}$, we are interested in the first-reaction time PMF, namely the probability distribution of its first-return (or passage) to the origin while in the reactive state. 

The conventional way in which one would approach such a problem is by writing and solving the Master equation for $C_{i}(x,n)$, the  probability that the boundary is at state $i\in\{\text{R},\text{NR}\}$, and the particle is at position $x$, after $n$ steps. The master equation for the bulk lattice sites reads

\begin{equation} \label{eq:32}
\begin{array}{ll}
  C_{\text{R}}(x,n+1) = \frac{1-p}{2} [  C_{\text{R}}(x+1,n) +  C_{\text{R}}(x-1,n)]  \\ \\+ \frac{q}{2}  [  C_{\text{NR}}(x+1,n) +  C_{\text{NR}}(x-1,n)],
\\  
\\
  C_{\text{NR}}(x,n+1) =\frac{1-q}{2} [  C_{\text{NR}}(x+1,n) +  C_{\text{NR}}(x-1,n)]  \\ \\+ \frac{p}{2} [  C_{\text{R}}(x+1,n) +  C_{\text{R}}(x-1,n)],  
\end{array}
\end{equation}
 
\noindent while the equations for the boundary ($x=0$) and near boundary ($x=1$) sites are given by 

\begin{equation} \label{eq:33}
\begin{cases}
C_{\text{R}}(0,n+1)=0,
\\ \\
C_{\text{R}}(1,n+1)=\frac{1-p}{2}C_{\text{R}}(2,n)+q[C_{\text{NR}}(0,n)+\frac{1}{2}C_{\text{NR}}(2,n)],
\\ \\
C_{\text{NR}}(0,n+1)=\frac{1-q}{2}C_{\text{NR}}(1,n)+\frac{p}{2}C_{\text{R}}(1,n),
\\ \\
C_{\text{NR}}(1,n+1)=(1-q)[C_{\text{NR}}(0,n)+\frac{1}{2}C_{\text{NR}}(2,n)]+\frac{p}{2}C_{\text{R}}(2,n).
\end{cases} 
\end{equation}

\noindent These equations are supplemented by an initial condition. For example, if we assume the particle starts at the boundary ($x_{0}=0$) in the non-reactive state, we have $C_{\text{R}}(x,0)=0$ and $C_{\text{NR}}(x,0)=\delta_{x,0}$. 

Although doable, solving the above set of equations brute force is a laborious task. To demonstrate this, we present a solution for the symmetric case,  $p=q$, in Appendix B. In the following, we instead note that the problem of a gated boundary is completely equivalent to the problem of gated particle switching between a reactive state and non-reactive state. We can thus evoke the results derived in Sec. II, and calculate the generating function of the first-reaction time by a simple ``plug and play" method. 

\begin{figure}[]
\begin{centering}
\includegraphics[width=0.9\linewidth]{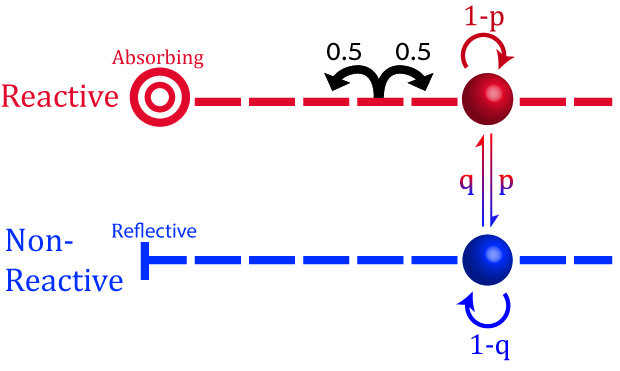}
\caption{Discrete-time symmetric random-walk on a semi-infinite interval with a gated boundary. At every time step the particle makes a jump left or right with equal probabilities. In addition, the boundary evolves, changing from being absorbing to reflecting (and vice versa), according to the transition probabilities in the figure.}
\end{centering}
\end{figure}

\subsection*{Applying the Renewal Approach}

\noindent To calculate the gated reaction time using Eq. (\ref{eq:9}), we require the Z-transform of the ungated first return-time $N_{FR}$. This is known to be given by \cite{klafter2011first}

\begin{equation}  \label{eq:34}
\tilde{N}_{\textrm{FR}}(z)=\sum_{n} f_{0}(n)z^n=1-\sqrt{1-z^2},
\end{equation}

\noindent where 

\begin{equation}  \label{eq:34_1}
f_{0}(n)=\begin{cases}
 \frac{1}{2^{n}(n-1)}{n \choose n/2}, \text{\hspace{2ex} Even n,}\\
0, \text{\hspace{11.7ex} Odd n,}
\end{cases}
\end{equation}

\noindent is the probability that a symmetric random walker which started at the origin returns to the origin for the first time at the n-th step.

Substituting Eq. (\ref{eq:34}) into Eq. (\ref{eq:9}), we immediately obtain the Z-transform of the gated reaction time 

\begin{equation}  \label{eq:35}
 \tilde{N}(\Vec{0}_{\textrm{NR}},z)=  \frac{\sqrt{1-(z\Delta)^2} - \sqrt{1-z^2} } {\sqrt{1-(z\Delta)^2} + K_{eq}\sqrt{1-z^2}}.
\end{equation}

In appendix C, we show how Eq. (\ref{eq:35}) can be inverted to find $f_{0,\text{NR}}(n)$, the probability that a reaction occurs exactly after $n$ steps where the walker starts on the boundary in the non-reactive state. In the symmetric case, $K_{eq}=1$, we find

\begin{equation} \label{eq:36_1}
f_{0,\text{NR}}(n) = \sum_{k=0}^{n} g(k) h(n-k), 
\end{equation}

\noindent where

\begin{equation}
h(n)=\begin{cases}
 \frac{1}{2^{n}(n-1)}{n \choose \frac{n}{2}}[1-\Delta^{n}], \text{\hspace{1ex} Even n,}\\
0, \text{\hspace{16.5ex} Odd n,}
\end{cases}
\end{equation}

\noindent and

\begin{equation}
g(n)=\frac{1}{16p(1-p)}\begin{cases}
 \frac{1}{2^{n}(n+1)}{n+2 \choose \frac{n}{2}+1}[1-\Delta^{n+2}], \text{\hspace{1ex} Even n,}\\
0, \text{\hspace{20.75ex} Odd n.}
\end{cases}
\end{equation}

\begin{figure*}[t]
\begin{centering}
\includegraphics[width=0.88\linewidth]{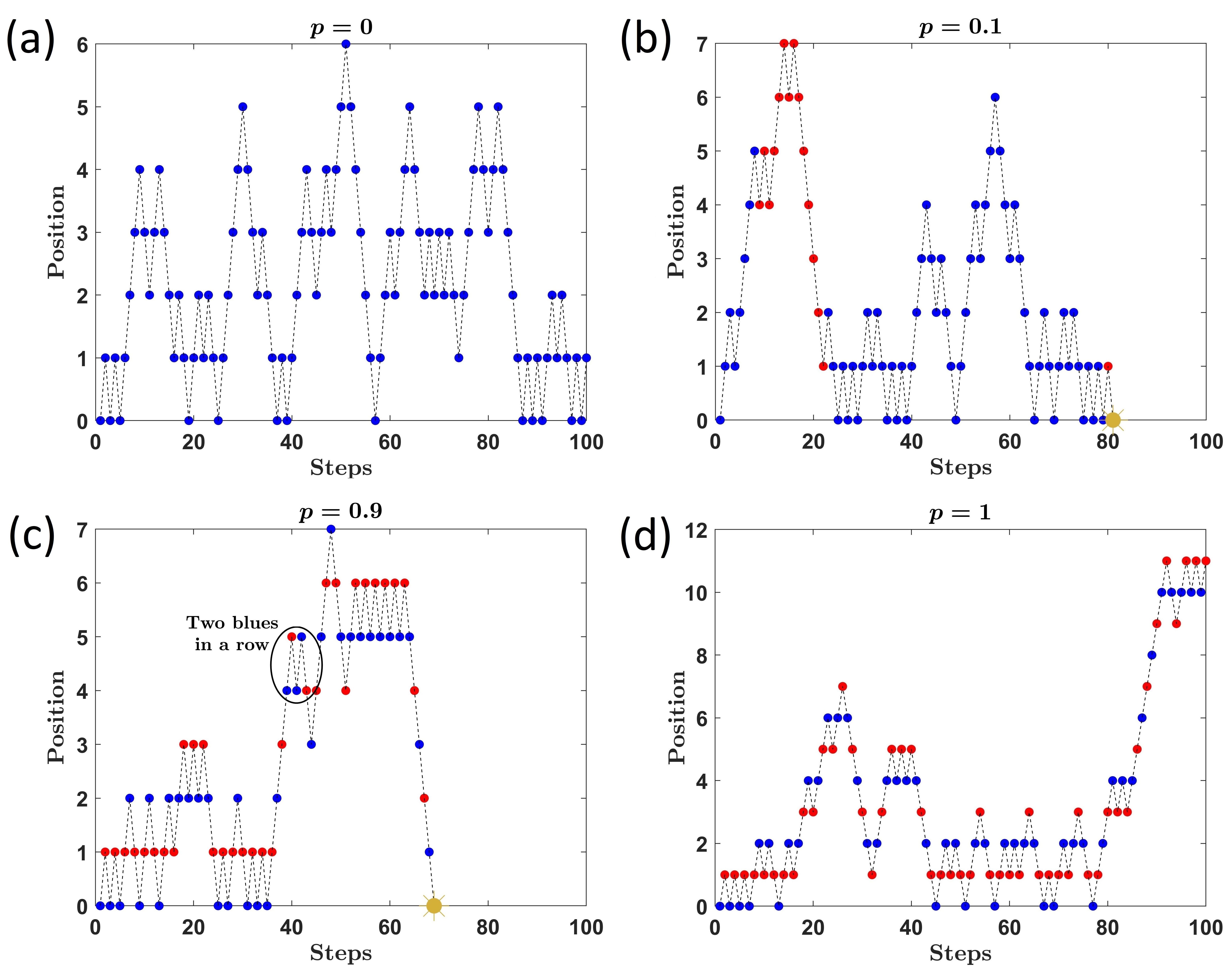}
\caption{Realizations of the gated reaction from Fig. 6 with symmetric internal dynamics ($p=q$), for $p=0,0.1,0.9$ and $1$ (a, b, c and d respectively). Blue (red) circles symbolize the walker being in a non-reactive (reactive) state. Gold circles symbolize reactions. Here, the particle starts at $x_{0}=0$ in the non-reactive state. It can be observed that for $p=0$ and $p=1$ a reaction cannot occur.}
\end{centering}
\end{figure*}

\noindent Exact formulas for the case $K_{eq} \neq 1$ are given in Appendix C. 

Equation (\ref{eq:36_1}) gives the full distribution of the gated reaction time, but since it is given in the form of a convolution it is hard to extract insight from it directly and without further analysis (one might say it is convoluted). Of main interest is the long time asymptotics, which can instead be readily obtained by utilizing the fact that the gated asymptotics inherits the ungated asymptotics.

For the ungated problem, by use of the Stirling approximation in Eq. (\ref{eq:34_1}), an asymptotic behaviour $f_{0}(n)\simeq \sqrt{\frac{2}{\pi}} n^{-3/2}$ is obtained for the even step sequence, while the probability of return on an odd step is zero. Averaging subsequent even and odd steps we get an average asymptotic behaviour of $f_{0}(n)\sim \frac{1}{\sqrt{2\pi}} n^{-3/2}$.\cite{klafter2011first} Note that the mean of the first return-time $N_{FR}$ diverges, and that the same is true for all its higher order moments.
To obtain the gated problem asymptotics, all we need to do is identify $C=\frac{1}{\sqrt{2 \pi}}$ and $\gamma=3/2$ in Eq. (\ref{eq:24}) and to substitute them, along with $\tilde{N}_{FR}(\Delta) = 1-\sqrt{1-\Delta^2}$, into Eq. (\ref{eq:27}). We then get, without any further calculation, that the average asymptotic behaviour for the gated problem is  

\begin{equation}  \label{eq:37}
f_{0,\text{NR}}(n) \simeq \frac{\nu}{\sqrt{2\pi}}n^{-3/2},
\end{equation} 

\noindent where we have defined

\begin{equation}   \label{eq:38}
\nu \equiv \frac{\pi^{-1}_{\textrm{R}} }{\sqrt{1-\Delta^2}}.
\end{equation}

\noindent As expected Eq. (\ref{eq:37}) asserts that the mean of the gated reaction time diverges along with all its higher moments. 

\subsection*{The Case of Symmetric Internal Dynamics}

\noindent In Fig. 7, we provide some realizations of the gated reaction depicted in Fig. 6. We focus on the symmetric case $p=q$. It can be seen that for $p=0$ and $p=1$ a reaction will never occur, albeit for completely different reasons. In the former case, the probability to transition to the reactive state is zero, and since the walker starts in the non-reactive state it simply stays in it. In the latter case, the walker is sure to switch internal states with every step it takes, such that it is in the reactive state for odd number of steps and in the non-reactive state for even number of steps. Furthermore, it is generally true that the walker can return to the origin only after taking an even number of steps. Thus the walker will never arrive there in the reactive state. We thus conclude that for $p \in \{0,1\}$ we have $f_{0,\text{NR}}(n)=0$, and equivalently that the survival probability is $S(n)=1$, for all $n$.

To better understand what happens for $p \in (0,1)$ we set $p=q$, in Eq. (\ref{eq:37}) to obtain (N.B. averaged subsequent even and odd steps)
\begin{equation}  \label{eq:40}
f_{0,\text{NR}}(n) \simeq
\frac{1}{\sqrt{2\pi p (1-p)}}n^{-3/2},
\end{equation} 

\noindent which corresponds to a survival probability of 
\begin{equation}  \label{eq:41}
S(n) \simeq \sqrt{\frac{2}{\pi p(1-p)}} n^{-\frac{1}{2}}.
\end{equation}

\noindent The asymptotics of Eq. (\ref{eq:40}) is corroborated in Fig. 8 for even $n$. To do so we  multiply the averaged result in Eq. (\ref{eq:40}) by a factor of $2$, since $f_{0,\text{NR}}(n)=0$ for odd  $n$. 

\begin{figure}[t!]
\includegraphics[width=0.9\linewidth]{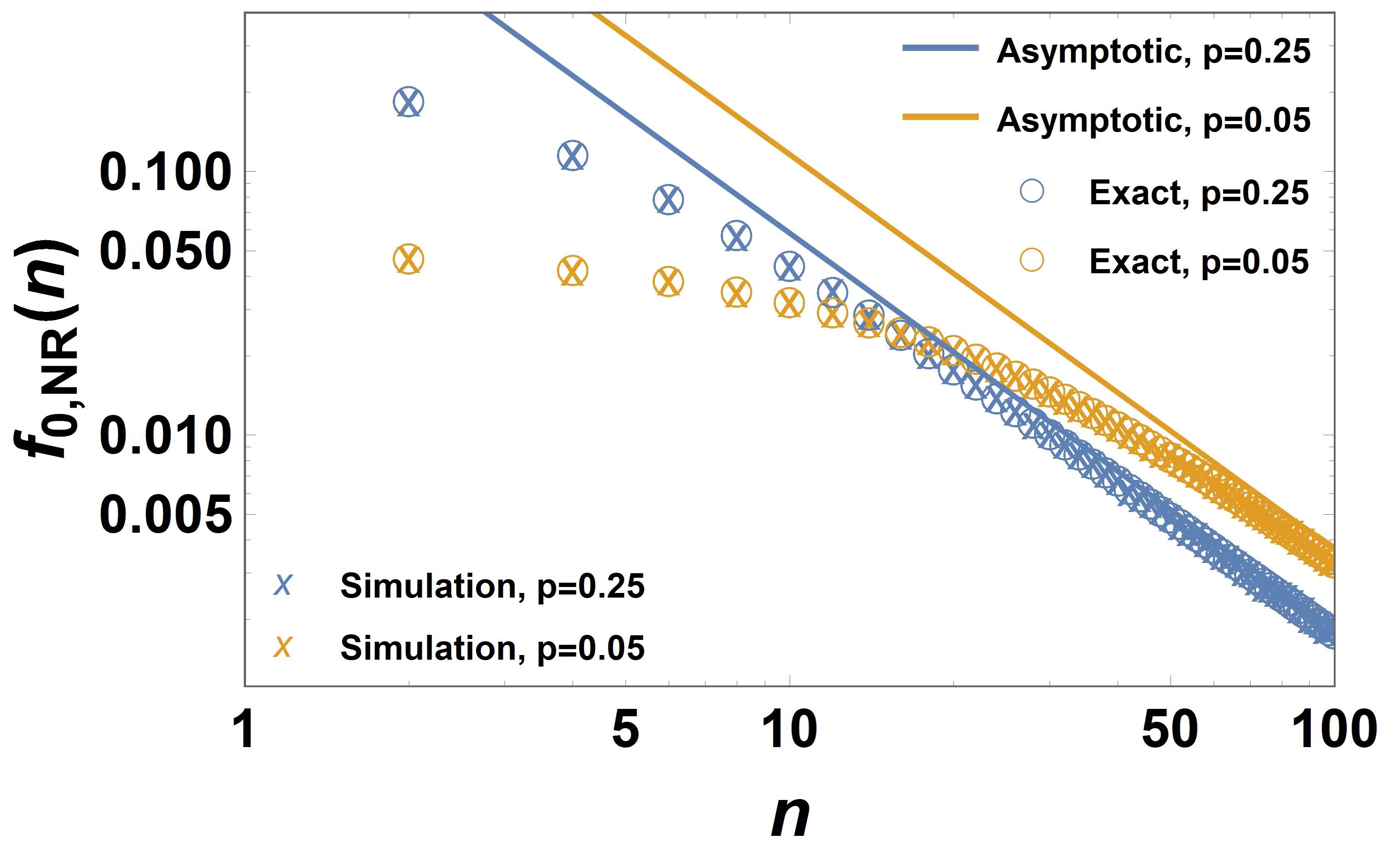}
\caption{A log-log plot of the PMF of the gated reaction time, $f_{0,\text{NR}}(n)$, vs. the number of steps $n$. Here, the scenario considered is identical to that in Fig. 7, and two values of $p=q$ are considered. The color coded open circles are exact analytical results coming from Eq. (\ref{eq:36_1}), and the X's come from  simulations ($10^6$ walkers were simulated). Lines are the respective asymptotics of Eq. (\ref{eq:40}), but with an added factor of 2 which takes care of parity (see  main text).}
\end{figure}
 
It is evident from Eqs. (\ref{eq:40}) and (\ref{eq:41}) that there is a complete symmetry about $p=0.5$ in terms of the first-passage and the survival probabilities. Indeed, in Sec. II we have derived a general expression for $\tilde{N}(\Vec{0}_{\textrm{NR}},z)$ for the case of symmetric internal dynamics, given by Eq. (\ref{eq:10}). In a short discussion below that equation, we have shown that when return to the origin is possible only on an even number of steps, the reaction time is invariant under the transformation $p \to 1-p$. The case under consideration here fulfills this criteria, and one can easily verify that this symmetry holds for all $n$. This is illustrated in Fig. 9a, where we calculated the reaction time probability after $n=2$, $20$, $50$ and $100$ steps for different values of $p$. Interestingly, the reaction time probability function obtains two maxima at optimal transition probabilities $p^{*}$ and $1-p^{*}$ which depend on $n$. As we explain below, this feature arises due to a certain resonance between the internal gating dynamics and the spatial dynamics which governs molecular encounters of the particle with the target at the origin.

\begin{figure*}[t]
\begin{centering}
\includegraphics[width=1\linewidth]{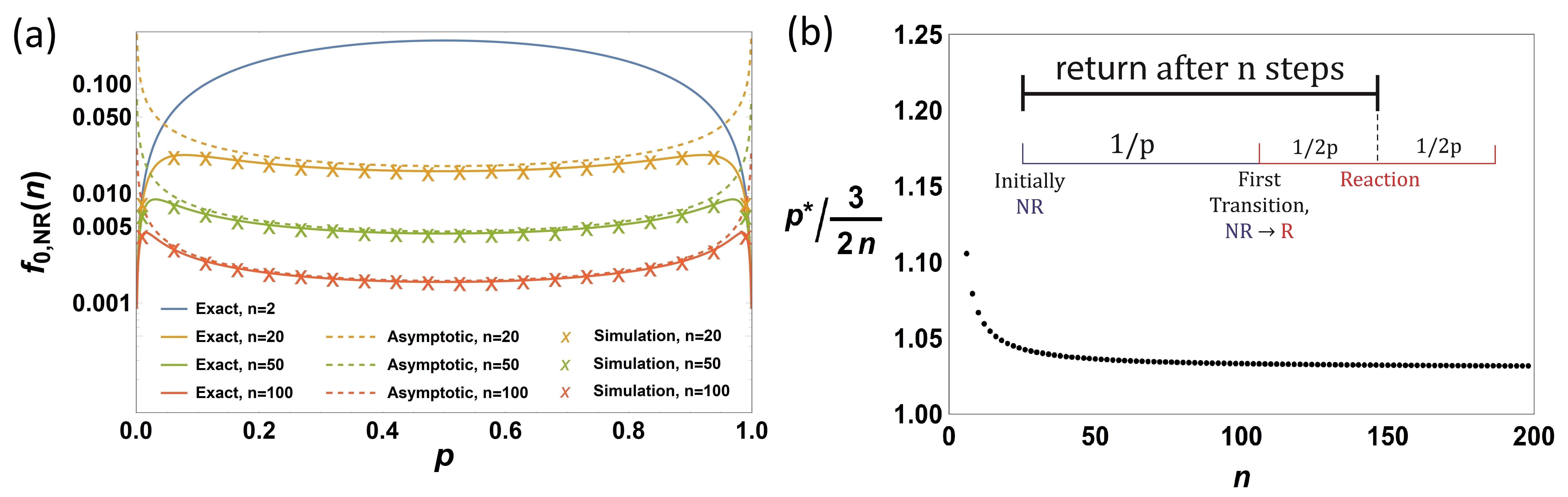}
\caption{(a) A log plot of the PMF of the gated reaction time $f_{0,\text{NR}}(n)$ for the symmetric case ($p=q$). Here, the number of steps was scanned across four fixed values $n=\{2,~20,~50,~100\}$ and plots were made vs. the transition probability $p$. The solid lines are exact analytical results coming from Eq. (\ref{eq:36_1}), dashed lines are the large $n$ asymptotics coming from Eq. (\ref{eq:40}), and X's come from  simulations ($10^6$ walkers were simulated for each case). Note the symmetry about $p=0.5$ and the two limiting cases of $p=0$ and $p=1$, in which $f_{0,\text{NR}}(n)=0$ for all $n$. The reaction time probability function obtains two maxima at optimal transition probabilities $p^{*}$ and $1-p^{*}$ which depend on $n$. (b) The plot shows that $p^{*} \simeq 3/2n$, which for large $n$ is accurate to within $\sim 3 \%$. The inset illustrates why this approximation works so well. In case the particle returns to the origin after $n$ steps, taking $p=3/2n$ would (on average) place the return time in  the middle of the particle's first reactive period.}
\end{centering}
\end{figure*}

As $n$ becomes large, $p^{*}$ tends to zero and the existence of its corresponding maximum becomes more pronounced. Note that the curvature is very different to the left and right of the optimal transition probability $p^{*}$. To the left  there is a rapid drop in reaction probability. To the right, the decrease in reaction probability is quite moderate. Of course, a mirror image is obtained for the maximum point at $1-p^{*}$, thus creating a ``safe zone" around $p=0.5$. While transition probabilities in this zone are not optimal, they still provide relatively high values for the reaction probability regardless of the specific value of $n$. This is interesting as it highlights an inherent trade-off between point-optimization and robustness in our problem. If one ever finds applications of this model in biological systems, it would be interesting to check whether evolution tuned the system to work with $p^{*}$ where small deviations from the optimum could decimate the probability to react, or rather in the safe zone which is fairly robust to changes in parameters.

We find that $p^{*}$ is inversely proportional to $n$. The exact proportionality factor can easily be determined by numeric maximization of Eq. (\ref{eq:36_1}), but let us give instead the following heuristic estimation. To increase the likelihood of a reaction occurring, we want to chose a transition probability $p$ that maximizes the probability of the particle to be reactive upon return to the origin. Focusing on situations where a return occurs exactly after $n$ steps, it makes sense to try and time things such that this return falls in the middle of a reactive period. As the transition probability is $p$, the average length of a reactive period is $1/p$; and one would want to maximize this duration to ensure that the reactive period covers the return time despite stochastic fluctuations. Naively, this could be done by taking $p$ to zero. However, recall that the particle starts in the non-reactive state and that transitioning to the reactive state takes $1/q$ units of time on average. As we have assumed $p=q$, it is clear that $p=3/2n$ is the smallest transition probability for which the return on average occurs in the middle of a reactive period (Fig. 9b inset). In Fig. 9b, we show that this intuitive argument approximates $p^{*}$ to a surprising accuracy, leaving just a mere $\sim 3 \%$ systemic relative error for large $n$.

\subsection*{Additional Interesting Cases}
\noindent So far we have only discussed the intriguing symmetric case of $p=q$. This case eloquently reveals new features that are not observed in the analogous continuous-time problem, and that are in fact due to the discretization of time. However, as mentioned above, we have also solved the problem for any choice of $p$ and $q$, and we refer the reader to Appendix C for the inversion of Eq. (\ref{eq:35}). In previous continuous-time works it was shown that for a certain range of transition rates an interesting transient behaviour can be observed.\cite{mercado2019first,scher2020unifying} In this so called ``cryptic regime" the transition rate from the non-reactive state to the reactive state is low with respect to the rate of the reversed transition. Markedly, there is a prolonged transient regime governed by a $\sim t^{-1/2}$ power law before the known $\sim t^{-3/2}$ asymptotics kicks in. The analogous $\sim n^{-1/2}$ cryptic regime can also be observed here by taking $p \gg q$.

Indeed, to obtain the gated problem transient behaviour under high cripticity, all we need to do is identify $C=\frac{1}{\sqrt{2 \pi}}$ and $\gamma=3/2$ in Eq. (\ref{eq:24}) and to substitute them, along with $\tilde{N}_{FR}(\Delta) = 1-\sqrt{1-\Delta^2}$, into Eq. (\ref{eq:51}) to obtain

\begin{equation}  \label{eq:53}
  f_{0,\text{NR}}(n) \simeq  \frac{\sqrt{1-\Delta^2}} {K_{eq}\sqrt{2\pi}}n^{-\frac{1}{2}},
\end{equation}

\noindent  where we understand this result as an average of subsequent even and odd steps. Since $f_{0,\text{NR}}(n)=0$ for odd  $n$, the result for the even sequence is two times that given in Eq. (\ref{eq:53}).

Finally, note that we have only considered here the case in which the particle is initially on the target in the non-reactive state. However, plugging the result in Eq. (\ref{eq:35}) into Eq. (\ref{eq:13}) one can solve for any initial condition. As in the continuous-time analogue,\cite{scher2020unifying} a bi-modal reaction time distribution can be observed when the particle starts away from the target in the reactive state, and for the right choice of $p$ and $q$. We trace this effect to the existence of two populations of particles: those which reached the origin and reacted without ever switching to the non-reactive state, and those which switched to the non-reactive state prior to reaching. As the latter are blocked from reacting for $1/q$ steps on average, two distinct peaks are formed when this time period is taken to be much larger than the median return time to the origin. 
\subsection*{General Perspective on Gated 1D Random Walks and Reactions}

\noindent To put the example solved above in its right context, we offer a very brief review of previous works on gated 1D random walks on the semi-infinite line. Specifically, while a lot of work has been dedicated to the classical random walk in discrete space and time, it seems that the gated version of the corresponding first-passage and return problems was first solved here. That said, solutions to the gated problem in  continuous time and space and in continuous time and discrete space have been presented before, and in the following we will review these cases for comparison. 

\textit{Continuous time and space} --- 
In his work on coagulation, Smoluchowski's was the first to build a model where diffusive transport is the rate limiting step of the reaction kinetics. To find this so called diffusion-controlled reaction rate, Smoluchowski solved for the flux of particles, initially uniformly distributed in an infinite pool of particles, into a 3D spherical symmetric sink fixed to the origin.\cite{chandrasekhar1943stochastic}  Many great reviews have been written on Smoluchowski's model and its generalizations.\cite{noyes1961effects,calef1983diffusion,rice1985diffusion,weiss1986overview} 

Although here we consider the first-passage to the boundary of a single-particle diffusing on the semi-infinite line, the treatment of the boundary is essentially the same. The solution can be obtained by solving the diffusion equation with an absorbing boundary condition  $c(0,t)=0$, where the first-passage distribution through the origin is given by the flux to the origin $f(0,t)=D\frac{\partial c(x,t)}{\partial x} |_{x=0}$. The equation is supplemented with the delta initial condition $c(x,0)=\delta(x_{0})$. Note that here $x_{0}=0$ is the trivial case in which the particle is immediately absorbed. Solving for $x_{0} > 0$, one gets a heavy-tailed distribution $f_{abs}(0,t) = \frac{x_{0}}{\sqrt{4\pi D t^3}}e^{-x^2_{0}/4Dt}$, with an asymptotic behaviour of $\sim x_{0}t^{-3/2}$ in the long-time limit.\cite{redner2001guide} 

In their seminal paper on diffusion-controlled reaction rates, Collins and Kimball extented Smoluchowski's model to allow for the case in which the target is reactive only in a fraction of the visits, leading naturally to a radiating (partially absorbing) boundary condition $D\frac{\partial c(x,t)}{\partial x}|_{x=0}=\kappa c(0,t)$.\cite{collins1949diffusion} When $\kappa=0$ there is zero flux through the origin, i.e., a reflecting boundary. Similarly, if $\kappa$ is taken to be very large, the absorbing boundary condition is restored. The corresponding single-particle model was solved by Sano and Tachiya for several cases of practical interest.\cite{sano1979partially}
Surprisingly enough, only recently Pal, Castillo and Kundu conducted a comprehensive study on diffusion in finite and semi-infinite 1D intervals, where analytical expressions for the probability density of the particle displacement and its first-passage distribution were computed.\cite{pal2019motion} 

A subtle point to notice is that gating is implicit in the radiating boundary condition. Indeed, consider a gate for which the internal dynamics is much faster than the time between consecutive particle-target collisions. In this case,  the time it takes for the internal gating state to reach equilibrium can be neglected, and the probability of the system to be in a reactive state when a collision occurs is always the equilibrium fractional population $\pi_{\textrm{R}}$. Thus, the partially-absorbing case can be thought of as a special case of the gated problem.

Recently, the gated version of diffusion on the 1D infinite line was solved for the survival probability.\cite{mercado2019first} Notably, it was shown that when the transition rate from the non-reactive state to the reactive state is much lower with respect to reversed transition, there is an intermediate regime governed by a power law $\sim t^{-1/2}$, before the known $\sim t^{-3/2}$ asymptotic kicks in. In Sec.   II of work, when exploring transient behaviour in high crypticity, we showed that this result is not limited to 1D, and extends to general networks via Eq. (\ref{eq:51}). 

\textit{Continuous time and discrete space} --- Here the underlying process can be formulated using the framework of the CTRW.\cite{klafter2011first} In a similar manner of invoking the Tauberian theorem, one can show that the probability density of the first-passage time to the origin follows $f(x_{0},t)\sim \nu t^{-3/2}$, where $\nu$ is some pre-exponential that depends on the initial location and the hopping rate of the random-walker. The gated counterpart of this problem was solved.\cite{caceres1995theory,re1996survival,scher2020unifying} Curiously enough, it was shown that the ungated problem and its gated counterpart share the same power law in the asymptotic behaviour, while differing only in the pre-exponential factor (the gated pre-exponential is dependent also on the internal dynamics). Here and in a previous work\cite{scher2020unifying}, we have proved that this is a universal feature of CTRW and discrete time gating, which extends beyond 1D problems and onto general networks in arbitrary dimensions.

\section*{V. Summary}

\noindent Stochastic gating, i.e., the random switching of molecules between reactive and non-reactive states, poses a barrier to the understanding and proper description of reaction kinetics. To overcome this, we presented a unified approach to gated reactions in discrete time and space. We showed that one can always express the distribution of the gated reaction time in terms of ungated first-passage and return times.  Thus, instead of solving directly for the gated reaction time, which can be extremely difficult even for relatively straightforward scenarios, one can instead apply an indirect, but much simpler, solution algorithm. Given a network of interest, and the time-independent laws which govern stochastic motion on this network, act as follows to obtain the distribution of the gated reaction time:

\begin{itemize}
  \item Obtain the distribution of the \textit{ungated} first-passage time $N_{FP}(\Vec{r_{0}})$, i.e., the number of steps a particle takes to go from its starting position $\Vec{r}_{0}$ to a target whose position is set to the origin (by convention and without loss of generality).
  \item Obtain the distribution of the \textit{ungated} first-return time to the origin $N_{FR}$, i.e., the number of steps that a particle which starts at the origin takes to get back to the origin.
  \item Compute the generating functions of the \textit{ungated} first-passage and return times: $\tilde{N}_{FP}(\Vec{r}_{0}, z)=\Braket{z^{{N}_{FP}(\Vec{r_0})}}$ and $\tilde{N}_{\textrm{FR}}(z)=\Braket{z^{{N}_{FR}}}$ correspondingly. Note, that in many cases it is actually easier to obtain these functions directly, thus bypassing computation of the first-passage and return distributions. One can then skip the first two stages.
  \item Substitute $\tilde{N}_{\textrm{FR}}(z)$ into Eq. (\ref{eq:9}) to obtain the generating function of the \textit{gated} reaction time $N(\Vec{0}_{\textrm{NR}})$, i.e., the total time it takes the particle to react given that it started from the origin in the non-reactive state. 
  \item Substitute $\tilde{N}_{FP}(\Vec{r}_{0}, z)$, $\tilde{N}_{\textrm{FR}}(z)$, and Eq. (\ref{eq:9}) into Eq. (\ref{eq:13}) to obtain the generating function of the \textit{gated} reaction time $N(\Vec{x}_{0})$, i.e., the total number of steps it takes the particle to react given that it started from an arbitrary  initial condition $\Vec{x}_{0}=(\Vec{r}_{0}, \omega_{0})$, where   $\Vec{r}_{0}$ stands for the particle's initial position, and $ \omega_{0}$ for its initial internal state which can be reactive or non-reactive. 
\end{itemize}

\noindent As we have shown, the mean, variance, asymptotics and additional properties of the gated reaction time can all be extracted from the generating functions in Eqs. (\ref{eq:9}) and (\ref{eq:13}).

We stress that the gated reaction times in Eqs. (\ref{eq:9}) and (\ref{eq:13}) depend on the probabilities $p$ and $q$, which govern stochastic transitions between the reactive state and the non-reactive state as depicted in Fig. 2. This dependence is manifested not only via the internal equilibrium constant $K_{eq}=p/q$ and occupancies $\pi_{\textrm{R}}=q/(p+q)$ and $\pi_{\textrm{NR}}=p/(p+q)$, but also via $\Delta=1-p-q$ which governs the rate of internal-state equilibration by Eqs. (\ref{eq:1}) and (\ref{eq:100}). From a physical stand-point, this can be traced back to the fact that the probability to react upon a particle-target encounter is not constant, but rather time dependent. Indeed, replacing the probability to react with the equilibrium probability to be reactive is an approximation which fails to take into account correlations between consecutive particle-target encounters. Hence, the need for the fully fledged theory of gated reactions that was developed here and in previous work.\cite{scher2020unifying}

The implementation of the solution algorithm presented above clearly hinges upon prior knowledge or computation of the ungated first-passage and return times. Luckily, these are known analytically in cases of particular interest,\cite{redner2001guide,klafter2011first,rudnick2004elements} and newly made discoveries add to this knowledge routinely.\cite{Giuggioli2020Exact} Relating gated reaction times to their ungated counterparts, the framework developed herein extends our knowledge on the long studied topic of first-passage. Importantly, all the results obtained apply beyond the context of chemical reactions and could e.g., be used to analyze analogous scenarios which arise when considering search of stochastically gated targets \cite{mercado2021first,mercado2021search} and to the problem of first-detection with intermittent sensing.\cite{kumar2021first}\\

\noindent\section*{Acknowledgments}
The authors wish to thank Ofek Lauber Bonomo for commenting on early versions of this manuscript. The authors wish to thank an anonymous reviewer for suggesting a calculation that led to the addition of a subsection dedicated to the transient power-law behaviour in the limit of high crypticity. S.R. acknowledges support from the Israel Science Foundation (grant No. 394/19). 

\noindent\section*{Data Availability}
The data that supports the findings of this study are available within the Communication and its appendixes.

\clearpage
\onecolumngrid
\appendix
\renewcommand{\theequation}{A\arabic{equation}}
\renewcommand{\thefigure}{A\arabic{figure}}
\renewcommand{\thesection}{A\Roman{section}} 
\renewcommand{\bibnumfmt}[1]{[A1]}
\renewcommand{\citenumfont}[1]{A#1}

\section*{Appendix A -- Derivation of Eqs. (\ref{eq:1}) \& (\ref{eq:100})} \setcounter{equation}{0}
\noindent Based on the Markov chain model in Fig. 2, we define the transition probability matrix (the stochastic matrix):  

\begin{equation}
S=\begin{bmatrix} 1-p && q \\ p && 1-q \end{bmatrix}.
\end{equation}

\noindent Note that $v_{1}=\begin{bmatrix} \frac{q}{p} && 1 \end{bmatrix}^T$ and  $v_{2}=\begin{bmatrix} -1 && 1 \end{bmatrix}^T$ are Eigenvectors of the transition matrix, with Eigenvalues $\lambda_{1}=1$ and $\lambda_{2}= 1-p-q \equiv \Delta$, respectively.

The conditional probability distributions to be in each of the states after $n$ steps, given $ \omega_{0} = \text{NR}$, can be calculated by applying the transition matrix n times to $\begin{bmatrix} 0 && 1 \end{bmatrix}^T$, 

\begin{equation}
\begin{bmatrix} P(\text{R}, n \mid \text{NR})  \\ P(\text{NR}, n \mid \text{NR})  \end{bmatrix}= S^{n}\begin{bmatrix} 0  \\ 1  \end{bmatrix}=\frac{p}{p+q}S^{n}v_{1}+\frac{q}{p+q}S^{n}v_{2}.
\end{equation}
Hence, 
\begin{equation}
\begin{bmatrix} P(\text{R}, n \mid \text{NR})  \\ P(\text{NR}, n \mid \text{NR})  \end{bmatrix}=\frac{p}{p+q}\lambda_{1}^{n}v_{1}+\frac{q}{p+q}\lambda_{2}^{n}v_{2}=\frac{p}{p+q}v_{1}+\frac{q}{p+q}\Delta^{n}v_{2}= 
\begin{bmatrix} \pi_{\textrm{R}}(1-\Delta^{n}) \\ \pi_{\textrm{NR}}+\pi_{\textrm{R}}\Delta^{n}  \end{bmatrix},
\end{equation}

\noindent where $\pi_{\textrm{R}}=\frac{q}{p+q}$ and $\pi_{\textrm{NR}}=\frac{p}{p+q}$.

\noindent In the exact same manner we can consider the case $ \omega_{0} = \text{R}$, i.e., the initial distribution is described by $\begin{bmatrix} 1 && 0 \end{bmatrix}^T$, 

\begin{equation}
\begin{bmatrix} P(\text{R}, n \mid \text{R})  \\ P(\text{NR}, n \mid \text{R})  \end{bmatrix}= S^{n}\begin{bmatrix} 1  \\ 0  \end{bmatrix}=\frac{p}{p+q}S^{n}v_{1}-\frac{p}{p+q}S^{n}v_{2}.
\end{equation}
Hence, 
\begin{equation}
\begin{bmatrix} P(\text{R}, n \mid \text{R})  \\ P(\text{NR}, n \mid \text{R})  \end{bmatrix}=\frac{p}{p+q}\lambda_{1}^{n}v_{1}-\frac{p}{p+q}\lambda_{2}^{n}v_{2}=\frac{p}{p+q}v_{1}-\frac{p}{p+q}\Delta^{n}v_{2}= 
\begin{bmatrix} \pi_{\textrm{R}}+\pi_{\textrm{NR}}\Delta^{n} \\ \pi_{\textrm{NR}}(1-\Delta^{n})  \end{bmatrix}.
\end{equation}

\renewcommand{\theequation}{B\arabic{equation}}
\renewcommand{\thefigure}{B\arabic{figure}}
\renewcommand{\thesection}{B\Roman{section}} 
\renewcommand{\bibnumfmt}[1]{[B1]}
\renewcommand{\citenumfont}[1]{B#1}

\section*{Appendix B -- Direct Solution of Eq. (\ref{eq:32}) for the symmetric case}
\setcounter{equation}{0}

\noindent Here we denote the reactive state (R) by a subscript 1 and the non-reactive state (NR) by a subscript 2, where the transition probabilities between these states are described in Fig. 2. We also keep the definition $\Delta:=1-p-q$ of the main text. Later on we will assume $p=q$ in a carefully chosen place, which will greatly simplify the calculation.

\subsection*{Master Equations for the Bulk Sites}
\noindent The master equation for the bulk sites -- Eq. (\ref{eq:32}) of the main text -- is given by:

\begin{equation}
\left\{\begin{array}{ccc}
  C_{1}(x,n+1) = \frac{1-p}{2} [  C_{1}(x+1,n) +  C_{1}(x-1,n)]  + \frac{q}{2}  [  C_{2}(x+1,n) +  C_{2}(x-1,n)]
\\  
\\
  C_{2}(x,n+1) =\frac{1-q}{2} [  C_{2}(x+1,n) +  C_{2}(x-1,n)]  + \frac{p}{2} [  C_{1}(x+1,n) +  C_{1}(x-1,n)],  
\end{array}\right.
\end{equation}

\noindent where $C_{i}(x,n)$ is the the  probability that the boundary is at state $i\in\{\text{R},\text{NR}\}$, and the particle is at position $x$, after $n$ steps. Z-transforming we get

\begin{equation}\label{B2}
\left\{ \begin{array}{ccc}
 \sum_{n=0}^{\infty} C_{1}(x,n+1) z^{n} = \frac{1-p}{2} \sum_{n=0}^{\infty}[  C_{1}(x+1,n)z^n + C_{1}(x-1,n)z^n]  + \frac{q}{2}  \sum_{n=0}^{\infty}[  C_{2}(x+1,n)z^n +  C_{2}(x-1,n)z^n]
\\  
\\
  \sum_{n=0}^{\infty}C_{2}(x,n+1)z^n = \frac{1-q}{2} \sum_{n=0}^{\infty}[  C_{2}(x+1,n)z^n +  C_{2}(x-1,n)z^n]  + \frac{p}{2}  \sum_{n=0}^{\infty}[  C_{1}(x+1,n) +  C_{1}(x-1,n)z^n]. 
\end{array} \right. 
\end{equation}

\noindent Now we can simplify the left side, and by invoking the initial condition: $C_{\text{R}}(x,0)=0$ and $C_{\text{NR}}(x,0)=\delta_{x,0}$. Doing so, we get

\begin{equation}\label{B3}
\begin{cases}
     \sum_{n=0}^{\infty} C_{1}(x,n+1) z^{n}= z^{-1}\sum_{n=0}^{\infty} C_{1}(x,n+1) z^{n+1}=
     z^{-1}\sum_{n=1}^{\infty} C_{1}(x,n) z^{n}=z^{-1}(\tilde{C}_{1}(x)-C_{1}(x,0))=z^{-1}\tilde{C}_{1}(x)
     \\  
      \sum_{n=0}^{\infty} C_{2}(x,n+1) z^{n}= z^{-1}\sum_{n=0}^{\infty} C_{2}(x,n+1) z^{n+1}=
     z^{-1}\sum_{n=1}^{\infty} C_{2}(x,n) z^{n}=z^{-1}(\tilde{C}_{2}(x)-C_{2}(x,0))=z^{-1}(\tilde{C}_{2}(x)-\delta_{x,0}),
\end{cases}
\end{equation}

\noindent where $\tilde{C}_{i}(x)$ is the Z-transform of ${C}_{i}(x,n)$. Inserting Eq. (\ref{B3}) back into Eq. (\ref{B2})

\begin{equation}
\begin{cases}
 z^{-1}\tilde{C}_{1}(x) = \frac{1-p}{2} [  \tilde{C}_{1}(x+1) +  \tilde{C}_{1}(x-1)]  + \frac{q}{2} [  \tilde{C}_{2}(x+1) +  \tilde{C}_{2}(x-1)]
\\  
\\
  z^{-1}\tilde{C}_{2}(x) = \frac{1-q}{2} [  \tilde{C}_{2}(x+1) +  \tilde{C}_{2}(x-1)]  + \frac{p}{2}  [  \tilde{C}_{1}(x+1) +  \tilde{C}_{1}(x-1)],
\end{cases}
\end{equation}

\noindent where we observed that $\delta_{x,0}=0$ for all bulk sites. Adding, and subtracting, the two equations above we get 

\begin{equation}
\begin{cases}
\tilde{C}_{+}(x)= \frac{z}{2}[\tilde{C}_{+}(x+1)+\tilde{C}_{+}(x-1)]
\\  
\\
\tilde{C}_{-}(x)= \frac{z\Delta}{2}[\tilde{C}_{-}(x+1)+\tilde{C}_{-}(x-1)],
\end{cases} 
\end{equation}

\noindent where $\tilde{C}_{\pm}(x)=\tilde{C}_{1}(x) \pm \tilde{C}_{2}(x)$ and \textbf{we where we have assumed symmetric internal dynamics $p=q$, such that $\Delta=1-2p$}.
\noindent Guessing $C_{+}(x)=A\omega^x$:

\begin{equation}
A\omega^x=\frac{z}{2} A(\omega^{x+1}+\omega^{x-1}),
\end{equation}

\begin{equation}
\frac{z}{2}\omega^{2}-\omega+\frac{z}{2}=0.
\end{equation}

\noindent And the possible solution are

\begin{equation}
\omega_{1,2}=\frac{1\pm\sqrt{1-z^2}}{z}.
\end{equation}

\noindent But the $\omega_{1}$ solution diverges with $x$, so we retain only the $\omega_{2}$ solution:

\begin{equation}
\tilde{C}_{+}(x)=A(\frac{1-\sqrt{1-z^2}}{z})^{x}.
\end{equation}

\noindent Similarly, guessing $C_{-}(x)=B\omega^x$: 

\begin{equation}
B\omega^x=\frac{z\Delta}{2} B(\omega^{x+1}+\omega^{x-1}),
\end{equation}

\begin{equation}
\frac{z\Delta}{2}\omega^{2}-\omega+\frac{z\Delta}{2}=0,
\end{equation}

\begin{equation}
\omega_{1,2}=\frac{1\pm\sqrt{1-(z\Delta)^2}}{z\Delta}.
\end{equation}

\noindent Once again, the $\omega_{1}$ solution diverges with $x$, so we retain only the $\omega_{2}$ solution. Summarizing:

\begin{equation}
\begin{cases}
\tilde{C}_{+}(x)=A(\frac{1-\sqrt{1-z^2}}{z})^{x}
\\  
\\
\tilde{C}_{-}(x)=B(\frac{1-\sqrt{1-(z\Delta)^2}}{z\Delta})^{x}
\end{cases}
\end{equation}

\noindent So that 

\begin{equation}
\begin{cases}
\tilde{C}_{1}(x)=\frac{\tilde{C}_{+}(x)+\tilde{C}_{-}(x)}{2}=\frac{A}{2}(\frac{1-\sqrt{1-z^2}}{z})^{x}+\frac{B}{2}(\frac{1-\sqrt{1-(z\Delta)^2}}{z\Delta})^{x}
\\  
\\
\tilde{C}_{2}(x)=\frac{\tilde{C}_{+}(x)-\tilde{C}_{-}(x)}{2}=\frac{A}{2}(\frac{1-\sqrt{1-z^2}}{z})^{x}-\frac{B}{2}(\frac{1-\sqrt{1-(z\Delta)^2}}{z\Delta})^{x}
\end{cases} 
\end{equation}

\noindent Setting $p=q=1$ (then $\Delta=-1$), we know that $C_{1}(x,n)$ must vanish for all sites where $x$ is even, and that $C_{2}(x,n)$ must vanish for all sites where $x$ is odd. It follows that

\begin{equation}
    A=-B.
\end{equation}

\noindent Hence,

\begin{equation}
\begin{cases}
\tilde{C}_{1}(x)=\frac{A}{2}\Big[(\frac{1-\sqrt{1-z^2}}{z})^{x}-(\frac{1-\sqrt{1-(z\Delta)^2}}{z\Delta})^{x}\Big]
\\  
\\
\tilde{C}_{2}(x)=\frac{A}{2}\Big[(\frac{1-\sqrt{1-z^2}}{z})^{x}+(\frac{1-\sqrt{1-(z\Delta)^2}}{z\Delta})^{x}\Big].
\end{cases}
\end{equation}

\subsection*{Master Equations for the Boundary Sites}

\noindent Sites $x=0$ and $x=1$ are governed by different equations that reflect the boundary conditions. For the reactive state we have

\begin{equation}
\begin{cases}
C_{1}(0,n+1)=0,
\\  
\\
z^{-1}\tilde{C}_{1}(0)=0,
\end{cases} 
\end{equation}

\noindent and 

\begin{equation}\label{B18}
\begin{cases}
C_{1}(1,n+1)=\frac{1-p}{2}C_{1}(2,n)+q(C_{2}(0,n)+\frac{1}{2}C_{2}(2,n))
\\  
\\
z^{-1}\tilde{C}_{1}(1)=\frac{1-p}{2}\tilde{C}_{1}(2)+q(\tilde{C}_{2}(0)+\frac{1}{2}\tilde{C}_{2}(2)).
\end{cases} 
\end{equation}

\noindent Similarly, for the non-reactive state (N.B. the inital condition $C_{2}(x,0)=\delta_{x,0})$)

\begin{equation}\label{B19}
\begin{cases}
C_{2}(0,n+1)=\frac{1-q}{2}C_{2}(1,n)+\frac{p}{2}C_{1}(1,n)
\\  
\\
z^{-1}(\tilde{C}_{2}(0)-1)=\frac{1-q}{2}\tilde{C}_{2}(1)+\frac{p}{2}\tilde{C}_{1}(1),
\end{cases} 
\end{equation}

\noindent and

\begin{equation}\label{B20}
\begin{cases}
C_{2}(1,n+1)=(1-q)(C_{2}(0,n)+\frac{1}{2}C_{2}(2,n))+\frac{p}{2}C_{1}(2,n)
\\  
\\
z^{-1}\tilde{C}_{2}(1)=(1-q)(\tilde{C}_{2}(0)+\frac{1}{2}\tilde{C}_{2}(2))+\frac{p}{2}\tilde{C}_{1}(2).
\end{cases} 
\end{equation}
\subsection*{Overall Solution}
\noindent Combining Eqs. (\ref{B18})-(\ref{B20}) and rearranging we get

\begin{equation}\label{B21}
 \frac{\tilde{C}_{1}(1)}{qz}-\frac{\tilde{C}_{2}(1)}{(1-q)z}=(\frac{1-p}{2q}-\frac{p}{2(1-q)})\tilde{C}_{1}(2), 
\end{equation}

\noindent and

\begin{equation}\label{B22}
 (\frac{1}{pz}-\frac{qz}{2})\tilde{C}_{1}(1)-\frac{(1-q)z}{2}\tilde{C}_{2}(1)=\frac{1-p}{2q}\tilde{C}_{1}(2)+\frac{1}{2}\tilde{C}_{2}(2) + 1.
\end{equation}

\noindent By assuming $p=q$ we already know from the equations of the bulk the forms of $\tilde{C}_{1}(2)$ and $\tilde{C}_{2}(2)$ up to the factor $A(z)$. Thus, in Eqs. (\ref{B21}) and (\ref{B22}) there are 3 unknowns: $A$, $\tilde{C}_{1}(1)$, $\tilde{C}_{2}(1)$. For the 3rd equation we use the Z-transform of the master equation for site $x=2$ in the reactive state:

\begin{equation}\label{B23}
\begin{cases}
C_{1}(2,n+1)=\frac{1-p}{2}(C_{1}(1,n)+C_{1}(3,n))+\frac{q}{2}(C_{2}(1,n)+C_{2}(3,n))
\\  
\\
z^{-1}\tilde{C}_{1}(2)=\frac{1-p}{2}(\tilde{C}_{1}(1)+\tilde{C}_{1}(3))+\frac{q}{2}(\tilde{C}_{2}(1)+\tilde{C}_{2}(3)),
\end{cases} 
\end{equation}

\noindent where we know that $\tilde{C}_{1}(2)$, $\tilde{C}_{1}(3)$ and $\tilde{C}_{2}(3)$ are set by the bulk solutions. Hence, Eq. (\ref{B23}) is another equation for the unknowns $A$, $\tilde{C}_{1}(1)$ and $\tilde{C}_{2}(1)$. Overall, we have three equations for three unknowns, solving (N.B. $p=q$) we obtain:

\begin{equation}
    A=\frac{4}{\sqrt{1-z^2}+\sqrt{1-(\Delta z)^2}},
\end{equation}

\noindent the equations for sites $x \geq 1$:

\begin{equation}\label{B25}
\begin{cases}
\tilde{C}_{1}(x)=\frac{2}{\sqrt{1-z^2}+\sqrt{1-(\Delta z)^2}}\Big[(\frac{1-\sqrt{1-z^2}}{z})^{x}-(\frac{1-\sqrt{1-(z\Delta)^2}}{z\Delta})^{x}\Big]
\\  
\\
\tilde{C}_{2}(x)=\frac{2}{\sqrt{1-z^2}+\sqrt{1-(\Delta z)^2}}\Big[(\frac{1-\sqrt{1-z^2}}{z})^{x}+(\frac{1-\sqrt{1-(z\Delta)^2}}{z\Delta})^{x}\Big],
\end{cases}
\end{equation}

\noindent and for the boundaries (from Eq. (\ref{B19}))

\begin{equation}
    \tilde{C}_{2}(0)=\frac{2}{\sqrt{1-z^2}+\sqrt{1-(\Delta z)^2}}=\frac{A}{2}.
\end{equation}

\noindent Note that indeed as $z\to 0$ we get $\tilde{C}_{2}(0)\to 1$.

\subsection*{First-Passage PMF}
\noindent For $x=1$ Eq. (\ref{B25})  becomes:

\begin{equation}
\begin{cases}
\tilde{C}_{1}(1)=\frac{2}{\sqrt{1-z^2}+\sqrt{1-(\Delta z)^2}}\Big[(\frac{1-\sqrt{1-z^2}}{z})-(\frac{1-\sqrt{1-(z\Delta)^2}}{z\Delta})\Big] \equiv	\phi(z) \Big(\Phi(z) - \Phi(z \Delta)\Big)
\\  
\\
\tilde{C}_{2}(1)=\frac{2}{\sqrt{1-z^2}+\sqrt{1-(\Delta z)^2}}\Big[(\frac{1-\sqrt{1-z^2}}{z})+(\frac{1-\sqrt{1-(z\Delta)^2}}{z\Delta})\Big]  \equiv	\phi(z) \Big(\Phi(z) + \Phi(z \Delta)\Big).
\end{cases}
\end{equation}

\noindent We can rewrite $\phi(z)$ as 

\begin{equation}
   \phi (z) = \frac{2}{\sqrt{1-z^2}+\sqrt{1-(\Delta z)^2}} = 2 \frac{\sqrt{1-z^2} - \sqrt{1-(\Delta z)^2}}{(\Delta^2 -1 ) z^2} = - \frac{\sqrt{1-z^2} - \sqrt{1-(\Delta z)^2}}{2p(1-p) z^2}.
\end{equation}

\noindent Using Newton's generalized binomial theorem

\begin{equation}
    \phi(z) = -\frac{1}{2p(1-p)} \Big[  \sum_{k=1}^{\infty}  {\frac{1}{2} \choose k} (-1)^k z^{2k-2} - \sum_{k=1}^{\infty}  {\frac{1}{2} \choose k} (-1)^k \Delta^{2k} z^{2k-2} \Big].
\end{equation}

\noindent Expanding the binomial coefficient using the identity $\left(\begin{array}{c}
1 / 2 \\
k
\end{array}\right)=\left(\begin{array}{c}
2 k \\
k
\end{array}\right) \frac{(-1)^{k+1}}{2^{2 k}(2 k-1)}$ we get 

\begin{equation}
    \phi(z) = \frac{1}{2p(1-p)} \Big[  \sum_{k=1}^{\infty}  \frac{1}{2k-1} {2k \choose k} 2^{-2k}  z^{2k-2} - \sum_{k=1}^{\infty}  \frac{1}{2k-1} {2k \choose k} 2^{-2k} \Delta^{2k} z^{2k-2} \Big].
\end{equation}

\noindent Inverse Z-transform gives 

\begin{equation}
g(n) = \frac{1}{8p(1-p)} \begin{cases} 0, \hspace{30ex}  \textrm{Odd n,} \\
\frac{1}{n+1}{n+2 \choose \frac{n}{2}+1}2^{-n}\Big(1-\Delta^{n+2}\Big), \hspace{8.5ex} \textrm{Even n.}
\end{cases}
\end{equation}

\noindent Similarly, using Newton's generalized binomial theorem on $\Phi(z)$ we get

\begin{equation}
    \Phi(z) =  \sum_{k=1}^{\infty}  {\frac{1}{2} \choose k} (-1)^{k+1} z^{2k-1} =   \sum_{k=1}^{\infty} \frac{1}{2k-1} {2k \choose k} 2^{-2k}  z^{2k-1}.
\end{equation}

\noindent Inverse Z-transform gives

\begin{equation}
h(n) =\begin{cases} \frac{1}{2n} {n+1 \choose \frac{n+1}{2}}2^{-n} , \hspace{20ex}  \textrm{Odd n,} \\
0, \hspace{29.5ex} \textrm{Even n.}
\end{cases}
\end{equation}

\noindent Note that the inverse transform of $\Phi(\Delta z)$ is given by $\Delta^n h(n)$.

The inverse Z-transform of $\tilde{C}_{1}(1)$ and $\tilde{C}_{2}(1)$ can be calculated using the convolution theorem

\begin{equation}\label{B34}
\begin{cases}
    C_{1}(1,n) =  \sum_{k=0}^{n} g(k) h(n-k)\Big( 1 - \Delta^{n-k}  \Big) 
    \\
    \\
  C_{2}(1,n) =  \sum_{k=0}^{n} g(k) h(n-k) \Big( 1 + \Delta^{n-k}  \Big) .

\end{cases} 
\end{equation}

\noindent Note that if $n$ is even both $C_{1}(1,n)$ and $C_{2}(1,n)$ are zero, since starting at $n=0$ at $x=0$ there is no way of being at $x=1$ at even $n$. Now to get the first-passage PMF all we need to do is consider the probability to be at $x=1$ at $n-1$ and jump to $x=0$ while being, or switching into, the reactive state in the n-th step:

\begin{equation}\label{B35}
  f_{0,\text{NR}}(n) = \frac{1-p}{2} C_{1}(1,n-1) +  \frac{q}{2}  C_{2}(1,n-1).
\end{equation}

Setting $p=q=(1-\Delta)/2$ and substituting Eq. (\ref{B34}) into Eq. (\ref{B35}), we obtain 

\begin{equation}
  f_{0,\text{NR}}(n) = \frac{1+\Delta}{4} C_{1}(1,n-1) +  \frac{1- \Delta}{4}  C_{2}(1,n-1)= \frac{1}{4}\sum_{k=0}^{n-1} g(k) h(n-k-1)\Big( 2  - 2\Delta^{n-k}  \Big)=\frac{1}{2}\sum_{k=0}^{n} g(k) h(n-k-1)\Big( 1  - \Delta^{n-k}  \Big),
\end{equation}

\noindent which after some algebra can be shown equivalent to Eq. (\ref{eq:36_1}) in the main text.

\subsection*{First Passage Long-Time Asymptotic}

\noindent The Z-transform of the survival probability is given by 

\begin{equation}
\tilde{S}(z)=\tilde{C}_{+}(0) + A\sum^{\infty}_{i=1} \Big(\frac{1-\sqrt{1-z^2}}{z}\Big)^i = \tilde{C}_{+}(0) + A(\frac{1-\sqrt{1-z^2}}{z})\sum^{\infty}_{i=1} \Big(\frac{1-\sqrt{1-z^2}}{z}\Big)^{i-1}
=\tilde{C}_{+}(0) +  A\frac{(\frac{1-\sqrt{1-z^2}}{z})}{1-\frac{1-\sqrt{1-z^2}}{z}}.
\end{equation}

\noindent Plugging in $\tilde{C}_{+}(0)$ and $A$ we get

\begin{equation}
\tilde{S}(z)=2\frac{z\sqrt{1-(\Delta z)^2}-(1-\sqrt{1-z^2})(\sqrt{1-(\Delta z)^2}-2)}{(z-1+\sqrt{1-z^2})(\sqrt{1-z^2}+\sqrt{1-(\Delta z)^2})}.
\end{equation}

\noindent Taking the limit $z\to 1$ 

\begin{equation}
\tilde{S}(z) \simeq {\sqrt \frac{2}{ p(1-p)}} \frac{1}{\sqrt{1-z}}.
\end{equation}

\noindent Invoking the Tauberian Theorem, the survival probability is

\begin{equation}
S(n) \simeq \sqrt{\frac{2}{\pi p(1-p)}} n^{-\frac{1}{2}}.
\end{equation}

\noindent Then accordingly

\begin{equation}\label{B41}
f_{0,\text{NR}}(n) = -\frac{d S(n)}{d n} \simeq \frac{1}{\sqrt{2\pi p(1-p)}} n^{-\frac{3}{2}}
\end{equation}

\noindent Note that the asymptotic behaviour above gives an average between consecutive even and odd step. In fact the probability of returning on an odd step is zero, and the probability of returning on an even step is double of what is written in Eq. (\ref{B41}). This point was further discussed in the main text.

\renewcommand{\theequation}{C\arabic{equation}}
\renewcommand{\thefigure}{C\arabic{figure}}
\renewcommand{\thesection}{C\Roman{section}} 
\renewcommand{\bibnumfmt}[1]{[C1]}
\renewcommand{\citenumfont}[1]{C#1}

\section*{Appendix C -- Inversion of Eq. (\ref{eq:35}) in the main text} \setcounter{equation}{0}

\noindent The equation we want to invert is Eq. (\ref{eq:35}) of the main text:

\begin{equation}\label{C1}  
  \tilde{N}(\Vec{0}_{\textrm{NR}},z) =  \frac{\sqrt{1-(z\Delta)^2} - \sqrt{1-z^2} } {\sqrt{1-(z\Delta)^2} + K_{eq}\sqrt{1-z^2}}.
\end{equation}

\subsection{The case of $K_{eq} = 1$}

Note the $K_{eq} = 1$ implies $p=q$.
Our equation then simplifies to 

\begin{equation}  
 \tilde{N}(\Vec{0}_{\textrm{NR}},z) =  \frac{\sqrt{1-(z\Delta)^2} - \sqrt{1-z^2} } {\sqrt{1-(z\Delta)^2} + \sqrt{1-z^2}}.
\end{equation}

\noindent Let us denote

\begin{equation}
     \Phi(z) = \sqrt{1-z^2},
\end{equation}

\noindent and the denominator of Eq. (C2) as

\begin{equation}
   \psi(z) = \frac{1}{\sqrt{1-(\Delta z)^2}+\sqrt{1-z^2}} =  \frac{ \Phi( z) - \Phi( \Delta z) }{ (\Delta^2 - 1) z^2 } =  \frac{ \Phi( \Delta z) - \Phi( z)}{4p(1-p) z^2},
\end{equation}

\noindent where in the last equality we have used $p=q$. For completeness, let us denote the numerator of Eq. (C2) as

\begin{equation}
   \chi(z) = \Phi(\Delta z) - \Phi(  z) .
\end{equation}

\noindent Now note that if we can inverse Z-transform $\psi(z)$ and $\chi(z)$ independently, we will be able to inverse Z-transform Eq. (C2) by using the convolution theorem.

Using Newton's generalized binomial theorem

\begin{equation}
    \Phi(z)=\sum_{k=0}^{\infty}  {\frac{1}{2} \choose k} (-1)^k z^{2k},
\end{equation}

\noindent and using the identity $\left(\begin{array}{c}
1 / 2 \\
k
\end{array}\right)=\left(\begin{array}{c}
2 k \\
k
\end{array}\right) \frac{(-1)^{k+1}}{2^{2 k}(2 k-1)}$, we rearrange to obtain 

\begin{equation}
    \Phi(z)=\sum_{k=0}^{\infty} \frac{(-1)^{k+1}}{2^{2k} (2k-1)} {2k \choose k} (-1)^{k} z^{2k} = - \sum_{k=0}^{\infty} \frac{1}{2^{2k} (2k-1)} {2k \choose k} z^{2k}.
\end{equation}

\noindent We thus have 

\begin{equation}
   \chi(z) =  \sum_{k=1}^{\infty} \frac{1}{2^{2k} (2k-1)} {2k \choose k} [1-\Delta^{2k}]z^{2k}  .
\end{equation}

\noindent Inverse Z-transform gives

\begin{equation}\label{C9}
h(n)=\begin{cases}
 \frac{1}{2^{n}(n-1)}{n \choose \frac{n}{2}}[1-\Delta^{n}], \text{\hspace{1ex} Even n,}\\
0, \text{\hspace{16.5ex} Odd n.}
\end{cases}
\end{equation}

\noindent Similarly, we have

\begin{equation}
   \psi(z) =    \frac{1}{4p(1-p)} \sum_{k=1}^{\infty} \frac{1}{2^{2k} (2k-1)} {2k \choose k} [1-\Delta^{2k}]z^{2k-2} .
\end{equation}

\noindent Inverse Z-transform gives

\begin{equation}
g(n)=\frac{1}{16p(1-p)}\begin{cases}
 \frac{1}{2^{n}(n+1)}{n+2 \choose \frac{n}{2}+1}[1-\Delta^{n+2}], \text{\hspace{1ex} Even n,}\\
0, \text{\hspace{20.5ex} Odd n.}
\end{cases}
\end{equation}

\noindent By using the convolution theorem, we have the inverse of Eq. (C2)

\begin{equation}
f_{0,\text{NR}}(n) = \sum_{k=0}^{n} g(k) h(n-k) 
\end{equation}


\subsection{The case when $K_{eq} \neq 1$}

\noindent Let us multiply both numerator and denominator of Eq. (\ref{C1}) by $\sqrt{1-(z\Delta)^2} - K_{eq}\sqrt{1-z^2}$  to get 

\begin{equation}\label{C13}  
  \tilde{N}(\Vec{0}_{\textrm{NR}},z) =  \frac{\sqrt{1-(z\Delta)^2} - \sqrt{1-z^2} } {\sqrt{1-(z\Delta)^2} + K_{eq}\sqrt{1-z^2}}   \frac{\sqrt{1-(z\Delta)^2} - K_{eq}\sqrt{1-z^2} } {\sqrt{1-(z\Delta)^2} - K_{eq}\sqrt{1-z^2}}.
\end{equation}

\noindent Using the same notation for $ \chi(z) = \Phi(\Delta z) - \Phi(  z) $ as before and defining $ \eta(z) = \Phi(\Delta z) -  K_{eq} \Phi(  z) $ we re-write Eq. (\ref{C13}) as

\begin{equation}\label{C14}  
  \tilde{N}(\Vec{0}_{\textrm{NR}},z) =  \frac{\chi(z)  \eta(z)}{1-K^{2}_{eq} + z^2 (K^{2}_{eq}-\Delta^2)} \equiv  \chi(z)  \eta(z) \xi(z)
\end{equation}

\noindent Now note that if we can inverse Z-transform $\chi(z)$, $\eta(z)$ and $ \xi(z)$ independently, we will be able to inverse Z-transform Eq. (C14) by using the convolution theorem.

\noindent Well, we already know the inverse of $\chi(z)$ from Eq. (\ref{C9}), and the inverse of $\eta(z)$ can be calculated in the same manner to give

\begin{equation}
u(n)=\begin{cases}
 \frac{1}{2^{n}(n-1)}{n \choose \frac{n}{2}}[K_{eq}-\Delta^{n}], \text{\hspace{1ex} Even n,}\\
0, \text{\hspace{18.5ex} Odd n.}
\end{cases}
\end{equation}

\noindent We are left with the task of computing the inverse of $ \xi(z)$. Assuming $K_{eq} \neq 1$ we now do some algebraic manipulations:

\begin{equation}
     \xi(z) = \frac{1}{1 - K^{2}_{eq} + z^2 (K^{2}_{eq}- \Delta^2)} = \frac{1}{K^{2}_{eq}- \Delta^2}\frac{1}{z^2 - \frac{K^{2}_{eq}-1}{K^{2}_{eq}- \Delta^2}  }.
\end{equation}

\noindent Let us now denote $A=K^{2}_{eq} - \Delta^2$ and $B=\sqrt{\frac{K^{2}_{eq}-1}{K^{2}_{eq}- \Delta^2}}$. We can then write 

\begin{equation}
    \xi(z) = \frac{1}{A}\frac{1}{(z-B)(z+B)}= \frac{1}{2AB}(\frac{1}{z-B}-\frac{1}{z+B})= -\frac{1}{2AB^2}(\frac{1}{1-\frac{z}{B}}+\frac{1}{1+\frac{z}{B}}),
\end{equation}

\noindent and expanding the two terms in the parenthesis 

\begin{equation}
    \xi(z) = -\frac{1}{2AB^2} \sum_{k=0}^{n} ( \frac{1}{(-B)^{n}} +   \frac{1}{B^{n}} ) z^n =  -\frac{1}{2(K^{2}_{eq} -1)} \sum_{k=0}^{n}  \frac{1 + (-1)^n}{B^{n}}  z^n   ,
\end{equation}
 
\noindent and so its inverse is given by 

\begin{equation}
   v(n) = -\frac{1}{2(K^{2}_{eq} -1)} \frac{1 + (-1)^n}{B^{n}}.
\end{equation}

\noindent We are now in a position to invert Eq. (\ref{C14}), which is equivalent to Eq. (\ref{C1}) for $K_{eq} \neq 1$: 

\begin{equation}
f_{0,\text{NR}}(n) = \sum_{k=0}^{n} \sum_{j=0}^{n-k} h(k) u(j) v(n-k-j).
\end{equation}

\end{document}